\documentclass[twocolumn]{aastex62}
\usepackage{amsmath}
\usepackage{graphicx}
\usepackage[graphicx]{realboxes}
\usepackage{epsf}
\usepackage{color}
\usepackage{enumitem}
\usepackage{hyperref}
\usepackage{multirow}
\usepackage{tabularx} 
\defcitealias{CG2018}{CG18}
\hypersetup{
    colorlinks=true,
    linkcolor=blue,
    filecolor=magenta,      
    urlcolor=blue,
    pdftitle={Overleaf Example},
    pdfpagemode=FullScreen,
    }

\begin{document}
\title{A Comprehensive Study of Open Cluster Chemical Homogeneity \\ using APOGEE and Milky Way Mapper Abundances}
\author[0009-0005-0182-7186]{Amaya Sinha}
\affiliation{Department of Physics \& Astronomy,  University of Utah, 115 South 1400 East, Salt Lake City, UT 84112, USA}

\author[0000-0001-6761-9359]{Gail Zasowski}
\affiliation{Department of Physics \& Astronomy,  University of Utah, 115 South 1400 East, Salt Lake City, UT 84112, USA}

\author[0000-0002-0740-8346]{Peter Frinchaboy}
\affiliation{Department of Physics \& Astronomy, Texas Christian University, Fort Worth, TX 76129, USA}

\author[0000-0001-6476-0576]{Katia Cunha}
\affiliation{Steward Observatory, University of Arizona, Tucson, AZ 85721, USA}
\affiliation{Observatório Nacional, Rua General José Cristino, 77, 20921-400 São Cristóvão, Rio de Janeiro, RJ, Brazil}

\author[0000-0002-7883-5425]{Diogo Souto}
\affiliation{Departamento de F\'isica, Universidade Federal de Sergipe, Av. Marcelo Deda Chagas, Cep 49.107-230, S\~ao Crist\'ov\~ao, SE, Brazil}

\author[0000-0002-4818-7885]{Jamie Tayar}
\affiliation{Department of Astronomy, University of Florida, Bryant Space Science Center, Stadium Road, Gainesville, FL 32611, USA}

\author[0000-0002-3481-9052]{Keivan Stassun}
\affiliation{Department of Physics and Astronomy, Vanderbilt University, VU Station 1807, Nashville, TN 37235, USA}

\begin{@twocolumnfalse}
    \begin{abstract}
Stars in an open cluster are assumed to have formed from a broadly homogeneous distribution of gas, implying that they should be chemically homogeneous. Quantifying the level to which open clusters are chemically homogeneous can therefore tell us about ISM pollution and gas-mixing in progenitor molecular clouds. Using SDSS-V Milky Way Mapper and SDSS-IV APOGEE DR17 abundances, we test this assumption by quantifying intrinsic chemical scatter in up to 20 different chemical abundances across 26 Milky Way open clusters. 
We find that we can place 3$\sigma$ upper limits on open cluster homogeneity within 0.02~dex or less in the majority of elements, while for neutron capture elements, as well as those elements having weak lines, we place limits on their homogeneity within 0.2~dex. Finally, we find that giant stars in open clusters are $\sim$0.01~dex more homogeneous than a matched sample of field stars.
    \end{abstract}
\end{@twocolumnfalse}

\section{Introduction}

Immediately after the Big Bang, the only elements in the universe were hydrogen, helium, and trace lithium. It took the formation of stars and galaxies to populate the universe with the rest of the periodic table. Therefore understanding where and how stars produce and disperse heavy elements is essential to understanding the enrichment of the universe. However, many questions regarding the chemical enrichment of the universe still remain unanswered. Specifically, there is still much uncertainty on how well-mixed giant molecular clouds are or how heavy elements get from their production sites into stars \citep{gce3,gce2,gce4,weinberg}. Fortunately, with a few exceptions, the surface abundances of stars are fossil records of the gas composition from the molecular cloud in which they formed. As a result, we can use the present-day chemistry of stars to learn about the chemistry of the Milky Way in the past. 

In the age of large astronomical surveys such as GALAH \citep{galah0,galah1,galah2,galah3,galah4}, LAMOST \citep{lamost}, RAVE \citep{rave1,rave2}, APOGEE \citep{majewski2017}, and Gaia \citep{gaia,gaia_rvs}, we can now probe the chemistry of stars in the Milky Way on the scale of $\sim$ 0.1 dex or smaller in multiple elements across different nucleosynthetic families, allowing us to trace different chemical enrichment pathways. Furthermore, we can now study the chemistry of the Milky Way at multiple different scales, from the simplest population in conatal binaries \citep{binaries} to large populations of dispersed field stars \citep{ness2021}. 

Stars in an open cluster (OC) are assumed to have formed from a broadly homogeneous distribution of gas at the same time, implying that they should all have the same age and be at the same distance \citep{lada}. Using the chemistry of OC stars, we can infer the chemistry of the gas available at that point in the Milky Way's history, in particular within the thin disk. In the past, it has been suggested that using assumptions of chemical homogeneity from simple stellar populations like open clusters, it would be possible to reconstruct a dissolved cluster purely by its members' chemistry \citep{chemicaltagging}. This technique, known as chemical tagging, has been a strong motivator for studies of cluster chemistry.

While many studies support this assumption of OC chemical homogeneity \citep{desilva1,bovy,ness,cheng2021}, there has been work showing that at least some clusters are chemically inhomogeneous \citep{ness}. For example, \citet{geisler2012} argued that NGC~6791 may not be chemically homogeneous, due to the presence of a potential Na-O anti-correlation, a relationship most commonly found in globular clusters \citep{gratton2001}. This would be an exciting result as NGC~6791 already unique in the Milky Way, as both the most massive and the most metal rich open cluster. However other studies, such as \citet{cunha2015}, have shown that it is chemically homogeneous within measurement uncertainties. For a more detailed discussion on previous studies of open cluster chemical homogeneity, see Section \ref{sec:Disc}.

There are reasons why OCs could be heterogeneous in specific elements. Slow neutron capture element abundances, such as Sr, Ba, and Zr, can change over a star's lifetime as it enters the AGB phase of its evolution \citep{agb_s_process}. Dredge-up, which occurs in stars on the giant branch, causes the star's convective envelope to expand, and it eventually gets deep enough to pull CNO-cycled elements to the surface, thereby altering the surface abundances we measure \citep[e.g.;][]{cno1,cno2,cno3}. NGC~6705 is an interesting OC regarding this effect, as it has been observed to also be enhanced in Na due to dredge up \citep{Loaiza_Tacuri2023}.

The surface abundances of elements such as Mg can be affected by effects like mass transfer and atomic diffusion \citep{diffsuion,souto}. However, the latter only weakly impacts the upper giant branch. Lastly, mass transfer \citep[e.g.;][]{milliman2015,bastian2013,abate2013} and pollution events such as planetary engulfment \citep[e.g.;][]{pinnesoult2001,laughlin1997, carlberg2010} 
can also alter a star's surface abundances.

However, if an open cluster was measured to have nonzero chemical scatter even accounting for these factors, that could point to interesting and understudied physics that may have occurred during the formation of the OC. Simulations have shown that turbulent mixing during cloud assembly naturally produces a stellar abundance scatter that is $\sim$ 5 times smaller than that in the natal gas \citep{feng2014}; suggest that this mixing could explain the observed chemical homogeneity of stars forming from the same molecular cloud. This is supported by recent work by \citet{binod2024} who find open clusters in FIRE-2 simulations \citep{fire1, fire2,latte} to have chemical scatter within 0.02~dex on average.

However, chemical inhomogeneity in real clouds could be due to effects not fully captured by simulations, related to internal turbulence and gas mixing within the progenitor molecular cloud or pollution events such as core collapse supernovae (CCSNe) that occurred earlier in the cluster's lifetime \citep[e.g.;][]{krumholz,looney2006}.

Quantifying the level of chemical homogeneity in open clusters across a broad set of elements from various nucleosynthetic families would provide the basis for understanding the physics of early OC formation. 

This work aims to constrain the chemical homogeneity in a large set of abundances and clusters to disentangle the causes of those chemical variances. The structure of the paper is as follows: Section \ref{sec:Data} outlines the survey data, verification of the abundance uncertainties, and determination of the cluster membership. Section \ref{sec:Meth} details the methodology and calculation of the intrinsic scatter within each [X/Fe] across the final cluster sample. Section \ref{sec:Res} presents the results of our work, and Section \ref{sec:Disc} compares our results to previous findings.

\section{Data}
\label{sec:Data}

\subsection{SDSS}

\subsubsection{SDSS-V/MWM}
The abundances and radial velocities (RVs) we use are primarily drawn from the Milky Way Mapper (MWM; J.A.~Johnson, \emph{in prep}), a component of the fifth generation of the Sloan Digital Sky Survey \citep[SDSS-V;][J. Kollmeier, \emph{in prep}]{sdssv}. We use data from Internal Product Launch 3 (IPL-3), which will form the basis for SDSS DR19 (K.~Hawkins, \emph{in prep}). This dataset builds off of
the observing strategies and survey goals outlined in SDSS Data Release 18 and includes observations of over a million targets \citep{dr18}.

SDSS-V/MWM uses two telescopes: the Sloan Foundation Telescope at APO \citep{apo} and the duPont Telescope at LCO \citep{lascampanas}. Both are outfitted with nearly identical custom-built 300-fiber APOGEE spectrographs \citep{spectrograph}, which reach a resolution of $R \sim 22,500$, spanning the range of wavelengths between 1.51-1.70 $\mu$m. Unlike in SDSS-IV, which used a plug-plate system, SDSS-V now uses robotic fiber positioners \citep{fps}, which benefited from the adoption of a three-element corrector for the Sloan telescope at APO \citep{connector}. 

Within IPL-3, three different data pipelines were used to analyze the data taken from APO and LCO: 
The \emph{Payne} \citep{payne}, The \emph{Cannon} \citep{cannon}, and the APOGEE Stellar Parameters and Abundances Pipeline \citep[ASPCAP;][]{ASPCAP2016}. Both The \emph{Payne} and The \emph{Cannon} are label-transfer methods that determine stellar labels from spectra after being trained on a set of spectra with known labels. They are differentiated by the fact that The \emph{Cannon} is a data-driven model which requires no information about stellar models. Rather measurements from the \emph{Cannon} inherit information from the models from its training data. The \emph{Payne} incorporates physical models directly into its analysis. While these datasets are similar in many aspects, comparing the limits derived through each of them will provide a stronger constraint on the true homogeneity of the OCs in our sample.

\subsubsection{SDSS-IV/APOGEE}

In addition to IPL-3, we use abundances and RVs from the seventeenth and final data release \citep[DR17;][]{dr17} of SDSS-IV's \citep{blanton2017} Apache Point Observatory Galaxy Evolution Experiment \citep[APOGEE;][]{majewski2017}, which contains over 700,000 stars. The initial targeting strategy for APOGEE-1 and APOGEE-2 are outlined in \citet{apogee1} and \citet{apogee2}, respectively, and the final targeting for APOGEE-2 is outlined in \citet{apogee2N} for APOGEE-2 north, and \citet{apogee2S} for APOGEE-2S. 

The details for the APOGEE data reduction pipeline are described in \citet{nidever}, and the details for ASPCAP are found in \citet{ASPCAP2016}. The description for the updates to these pipelines for DR17 is included in Holtzman et al.\,(\emph{in prep}). The MARCS model atmospheres and interpolation methodology used in APOGEE are described by \citet{modelatmos} and \citet{MARCS}. The line lists used for DR17 are outlined in \citet{linelist}, and the spectral fitting used for ASPCAP is described in \citet{FERRE}. The details describing the APOGEE spectral grids can be found in \citet{NLTE}, \citet{Synspec}, and \citet{HubenySynspec}, and lastly, the details for Turbospectum can be found in \citet{TURBOSPECTRUM} and \citet{alvzrez_TURBO}.

We also include abundances from the BACCHUS Analysis of Weak Lines in APOGEE Spectra \citep[BAWLAS;][]{bawlas}, a value-added catalog (VAC) in DR17. This VAC provided abundances for several chemical species having weak and blended lines that cannot be reliably analyzed using ASPCAP. This sample consists of high signal to noise (SNR $>$ 150) red giant stars with no flags in either \texttt{STARFLAG} or \texttt{ASPCAPFLAG} and analyzed using the BACCHUS code \citep{bacchus}, which measures line-by-line elemental abundances from on-the-fly spectral synthesis. High quality measurements are stacked to create a sample of elemental abundances for elements with weak or blended lines. We use the BAWLAS VAC abundances and uncertainties for the following elements: Na, P, S, V, Cu, Ce, and Nd.

Two separate uncertainties are reported for each BAWLAS abundance measurement. One is the \texttt{X\_FE\_ERR\_MEAS} describing the measured uncertainty from the combined spectra using the same methodology as ASPCAP. The other is \texttt{X\_FE\_ERR\_EMP}, which describes the uncertainty derived from the spectral lines themselves. Here to remain consistent in our analysis we use \texttt{X\_FE\_ERR\_MEAS} as it is the closest to the uncertainty calculation method that we verify in Section \ref{sec:uncs}.

The calculation of the abundances in the BAWLAS catalog is outlined fully in \citet{bawlas}. Each spectrum has an associated \texttt{X\_SPECTRA\_FLAG}, with values \texttt{0}, \texttt{1}, or \texttt{9}. Both \texttt{0} and \texttt{9} indicate either suspicion with the final fit or total failure, and only \texttt{1} indicates that the spectral fit is trustworthy. To ensure the highest quality sample, we require all the stars used in this study to have measurements with {\texttt{X\_SPECTRA\_FLAG} = 1}.

\subsubsection{Quality Cuts}
\label{sec:quality_cuts}
We limit our sample in APOGEE DR17 to stars with \texttt{VERR~<}~0.1~km~s$^{-1}$ and \texttt{|VHELIO\_AVG|~<}~5000~km~s$^{-1}$ to ensure our stars have reliable radial velocities. We also restrict our sample to stars with \texttt{VSCATTER <} 1 km~s$^{-1}$, to remove potential binaries within our sample \citep{badenes2018,whelan2020}. Here \texttt{VHELIO\_AVG} refers to the average radial velocity derived from individual RVs that are drawn from cross-correlation of individual spectra with combined spectrum. \texttt{VERR} refers to the uncertainty on that radial velocity, and \texttt{VSCATTER} refers to the scatter of individual visit RVs around the average. To ensure the sample has reliable measurements we enforce a \texttt{SNR > 50}. These limits are identical between DR17 and IPL-3. We also limit our sample to stars between 3000~K and 6500~K, and $\rm -1 \leq [Fe/H] \leq 1$.

We also exclude from our sample any stars that have [X/Fe] flags in more than 2 elements. Lastly, we only sample from stars with \texttt{LOGG $\leq$ 3.5} in order to ensure that every star we study is a member of the giant branch. These requirements result in a sample of 305,201 stars. When using IPL-3 we use the corresponding columns and limits, with the exception of \texttt{VSCATTER} which is not included in IPL-3. 

In DR17's allStar file, we enforce quality cuts on the \texttt{STARFLAG} and \texttt{ASPCAPFLAG} columns, the details of which are included here: \url{https://data.sdss.org/datamodel/files/APOGEE_ASPCAP/APRED_VERS/ASPCAP_VERS/allStar.html}. The details of the APOGEE bitmasks are located here \url{https://www.sdss4.org/dr17/algorithms/bitmasks/}. Within the \texttt{ASPCAPFLAG} column we enforced the following requirements:
\begin{enumerate}
 \item \texttt{BITMASK 23; STAR\_BAD} == 0; BAD overall for star: set if any of \texttt{TEFF}, \texttt{LOGG}, \texttt{CHI2}, \texttt{COLORTE}, \texttt{ROTATION}, SN error are set, or any parameter is near grid edge (\texttt{GRIDEDGE\_BAD} is set in any \texttt{PARAMFLAG})
 \item \texttt{BITMASK 19; METALS\_BAD} == 0; FERRE failed to return a value for metals.
 \item \texttt{BITMASK 20; ALPHAFE\_BAD} == 0; Elemental abundance from window differs $>$ 0.5 dex from parameter abundance for [$\alpha$/Fe].
\end{enumerate}
and within the \texttt{STARFLAG} we made the following cuts:
\begin{enumerate}
 \item \texttt{BITMASK 2; BRIGHT\_NEIGHBOR} == 0; Star has neighbor more than 10 times brighter.
 \item \texttt{BITMASK 3; VERY\_BRIGHT\_NEIGHBOR} == 0; Star has neighbor more than 100 times brighter.
 \item \texttt{BITMASK 17; SUSPECT\_BROAD\_LINES} == 0; WARNING: cross-correlation peak with template significantly broader than autocorrelation of template.
\end{enumerate}

The IPL-3 ASPCAP allStar file does not have a published \texttt{ASPCAPFLAG} column. As such, we require that all the stars in our sample have \texttt{flag\_bad == False} and \texttt{flag\_warn == False}. For all three IPL-3 pipelines we use we enforce a \texttt{result\_flags == 0} requirement. Where possible we enforce a \texttt{x\_h\_flag == 0} requirement within each pipeline. Within the three IPL-3 allStar files we use, we enforce these quality cuts on the \texttt{spectrum\_flag} column, which has bits that correspond to DR17's \texttt{STARFLAG} column.

\subsection{Cluster Membership}
We start from the catalog of cluster members published in \citet{CG2018}, hereafter \citetalias{CG2018}, which contains membership information for over 200,000 stars across $\sim$ 2000 OCs using Gaia DR2 \citep{dr2}. Due to the more recent availability of Gaia DR3 \citep{edr3,dr3}, we first match the stars identified as cluster members in \citetalias{CG2018} to their DR3 kinematics and positions. \citetalias{CG2018} used two spatial (RA $\alpha$, DEC $\beta$) and two kinematic (proper motion-RA $\delta _{\alpha *}$, proper motion-DEC $\delta _{\beta *}$) parameters, as well as parallax $\varpi$, as inputs into an unsupervised machine learning algorithm to determine cluster membership. We limit the initial cluster membership candidacy to stars from \citetalias{CG2018} within three cluster radii of their cluster centers. Using each cluster's distribution in radial velocity, we find that a minimum probability cut of P $\geq$ 0.5 in the \citetalias{CG2018} catalog maximized the overlap in membership within DR17 and IPL-3 while also minimizing contamination from non-cluster members. Of the 2019 OCs identified in \citetalias{CG2018}, only 145 clusters have any members within both DR17 and IPL-3.

\subsubsection{Kinematic Selection}
\label{sec:kinematics}
\begin{figure}[h]
  \centering
  \includegraphics[width=0.4\textwidth]{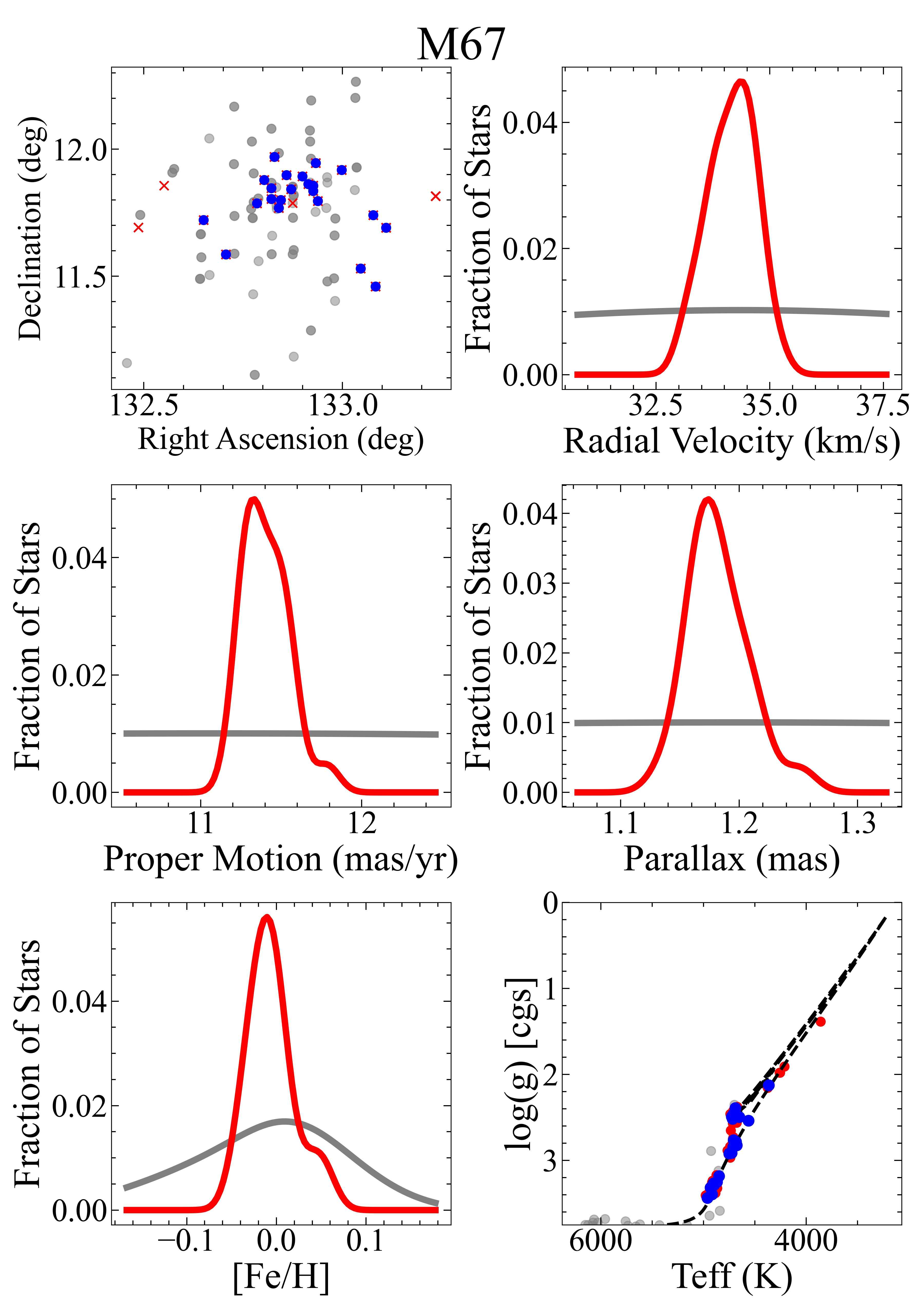}
  \caption{Cluster membership plot for M67. \emph{Top Left:} The distribution of stars on the sky in RA and DEC, with cluster members colored in blue, cluster member candidates that did not pass Section \ref{sec:kinematics}'s kinematic and parallax cuts in red, and field stars in grey. \emph{Top Right:} The KDE distribution of radial velocities of the cluster members compared to nearby field stars. \emph{Middle Left:} The distribution of the combined proper motion of the cluster members compared to nearby field stars. \emph{Middle Right:} The distribution of parallaxes of the cluster members compared to nearby field stars. \emph{Bottom Left:} The distribution of [Fe/H] of the cluster members compared to nearby field stars; note that this was not used to prune membership. \emph{Bottom Right:} The Kiel diagram of cluster members compared to surrounding stars. We also plot the MIST isochrone \citep{mist1,mist2} for this cluster's age and metallicity in black.
  }
  \label{fig:kinematic}
\end{figure}

To further ensure that the clusters identified had minimal contamination, we use a Kernel Density Estimator with a variable bandwidth, following Silverman's Rule, to measure the dispersion in four dimensions (radial velocity, $\delta _{\alpha *}$, $\delta _{\beta *}$, and $\varpi$). We reject stars further than two standard deviations from the cluster median. For an example of this methodology, see Figure \ref{fig:kinematic}, which shows the final distributions after these cuts in M67. The kinematic selection plots for all clusters in our sample can be found in Appendix \ref{sec:appendix_b}.

We plot the [Fe/H] distribution of each cluster against the distribution of nearby non-cluster members between 2--5 cluster radii, as published in CG2018. We visually inspect the HR diagrams for the cluster members compared to the annulus to ensure no contamination, as each cluster should follow a single distinct isochrone. We use MIST isochrones \citep{mist1,mist2}, generated using ages from \citet{cg2020}, and the median cluster [Fe/H]. Lastly, we only select clusters with $N \ge 6$ members, resulting in a final sample set of 26 open clusters. Determination of this minimum limit and the final sample size is related to material in Section \ref{sec:corrections}.

\begin{figure*}
 \centering
 \includegraphics[width=\textwidth]{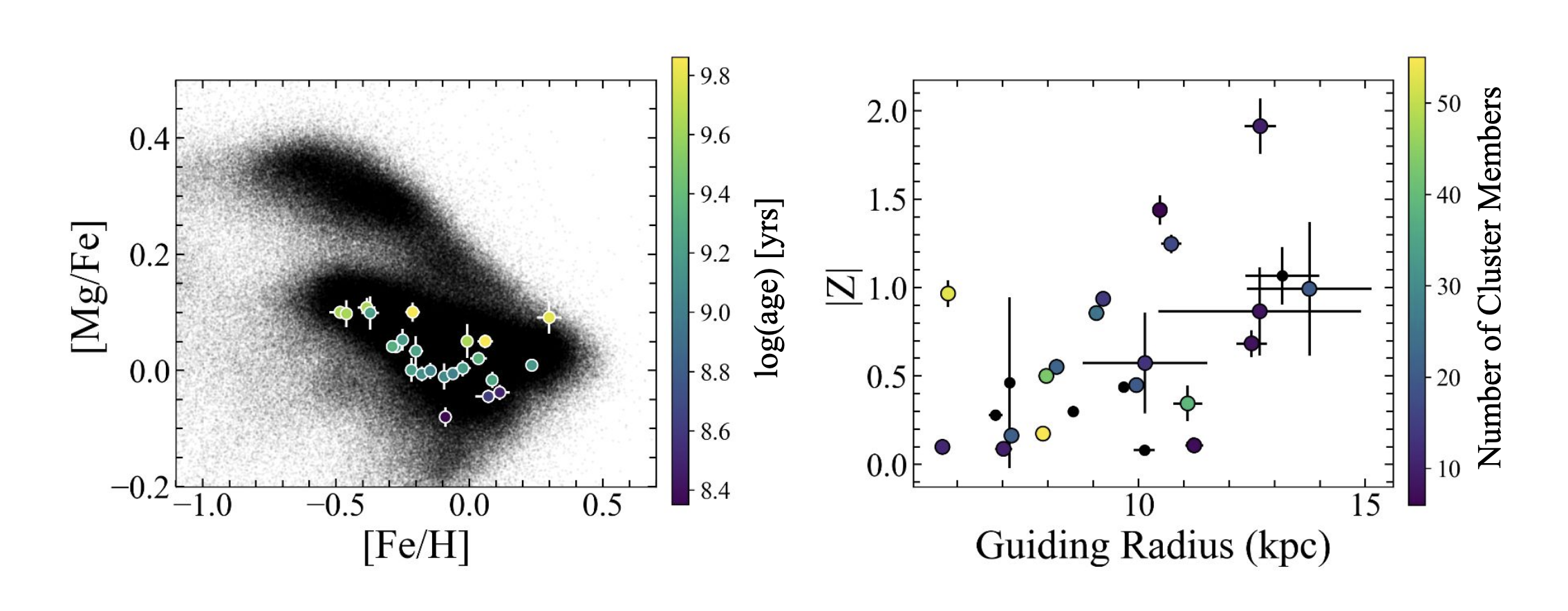}
 \caption{\emph{Left}: A plot of the [Mg/Fe] versus [Fe/H] plane}
 where the full giant star sample in APOGEE DR17 (subject to the same quality cuts as in Section~\ref{sec:quality_cuts}) is shown in black, and the clusters are colored by log(age). \emph{Right}: The guiding radius and galactocentric height 
 of all the clusters in our sample, colored by number of member stars as described in Section \ref{sec:kinematics}.
 \label{fig:cluster_params}
\end{figure*}

\subsubsection{Final Cluster Sample}
The distribution of the final cluster sample in radius, age, and [Mg/Fe] versus [Fe/H] can be seen in Figure \ref{fig:cluster_params}. 

We calculate the positional, kinematic, and orbital information for each cluster in our final cluster sample. Using Astropy SkyCoords \citep{astropy:2013, astropy:2018, astropy:2022} we calculate the Cartesian X, Y, Z galactocentric coordinates for all the clusters in our sample, as well as the galactocentric radius.

We integrate each cluster's orbit to measure its Z$_{max}$, eccentricity, and guiding radius using Galpy, with the \texttt{MWPotential2014} gravitational potential \citep{galpy1, galpy2} 
While this potential lacks a bar, no member of our sample is close enough to the Galactic center to produce a noticeable difference in the final measured eccentricity, guiding radius, or maximum height above the galactic plane. While there have been measured effects on these parameters due to the Milky Way's spiral arms, as seen in \citet{carrera2022}, as we do not use these parameters in the final science results, we do not include spiral arms in our potential.

We use a Monte Carlo method with N = 100 iterations to estimate uncertainties on these parameters. We measure 3D space velocity dispersion as a proxy for cluster mass and using the methodology outlined in \citet{weinberg}, we quantify the ratio of nucleosynthetic enrichment within each cluster from CCSNe and Type Ia supernovae. 

Lastly, we include age estimates on all our clusters from \citet{cg2020}. These ages were derived using an artificial neural network trained on a sample of reference clusters. For further description see \citet{cg2020}. At this stage we also flag red clump stars by eye within each cluster. All the cluster parameters are included in Table \ref{table1}, and will be included as a machine readable table.

\begin{table*}
\caption{List of columns in the table of cluster parameters.}
\begin{tabular}{l|l|l}
\label{table1}
 Column Name& Units &Column Description \\
 \hline
 \hline
 cluster & & Cluster name\\
 Nmems & & Number of identified cluster members identified in DR17 that passed all kinematic and quality cuts. \\
 RA & deg & Central right ascension for all the cluster members, from the median of Gaia DR3 coordinates\\
 DEC & deg & Central declination for all the cluster members, from the median of Gaia DR3 coordinates \\
 parallax & mas & Median parallax of cluster members from Gaia DR3\\
 parallax\_error & mas & Standard deviation parallax of cluster members from Gaia DR3 \\
 pmRA & mas~yr$^{-1}$ & Median right ascension proper motion of cluster members from Gaia DR3\\
 pmRA\_error & mas~yr$^{-1}$ & Standard deviation of right ascension proper motion of cluster members from Gaia DR3 \\
 pmDEC & mas~yr$^{-1}$ & Median declination proper motion of cluster members from Gaia DR3 \\
 pmDEC\_error& mas~yr$^{-1}$ & Standard deviation of declination proper motion of cluster members from Gaia DR3 \\
 RV&km~s$^{-1}$ & Median radial velocity of cluster members\\
 RV\_sigma&km~s$^{-1}$ & Standard deviation of radial velocity of cluster members \\
 RV\_sigma\_error&km~s$^{-1}$ & Uncertainty on RV\_sigma calculated by bootstrapping the cluster\\
 TV&km~s$^{-1}$ & Median tangential velocity, calculated by v$_{tangential}$ = 4.74$\mu$$\varpi^{-1}$, where $\mu$ is the total proper motion\\
 TV\_sigma&km~s$^{-1}$ & Standard deviation in tangential velocity of cluster members\\
 TV\_sigma\_error&km~s$^{-1}$ & Uncertainty on TV\_sigma calculated by bootstrapping the cluster\\
 SV&km~s$^{-1}$ & Median 3D space velocity calculated by v$_{3D}$ = $\sqrt{v_{radial}^{2} + v_{tangential}^{2}}$\\
 SV\_sigma&km~s$^{-1}$ & Standard deviation in 3D space velocity of cluster members \\
 SV\_sigma\_error&km~s$^{-1}$ & Uncertainty on SV\_sigma calculated by bootstrapping the cluster \\
 Rgc& kpc & Median galactocentric radius from Galpy integration\\
 Rgc\_error&kpc & Uncertainty on galactocentric radius from Galpy integration\\
 Rguiding&kpc & Median guiding radius from Galpy integration\\
 Rguiding\_error&kpc & Uncertainty on guiding radius from Galpy integration\\
 Zmax&kpc & Median maximum distance from galactic midplane from Galpy integration\\
 Zmax\_error&kpc & Uncertainty on the maximum distance from galactic midplane from Galpy integration\\
 e& & Median eccentricity from Galpy integration\\
 e\_error& & Uncertainty on eccentricity from Galpy integration\\
 X&kpc& Median x galactocentric coordinate using Cartesian coordinates \\
 X\_error&kpc & Uncertainty on x galactocentric distance using Cartesian coordinates \\
 Y&kpc & Median y galactocentric coordinate using Cartesian coordinates\\
 Y\_error&kpc & Uncertainty on y galactocentric coordinate using Cartesian coordinates\\
 Z&kpc & Median z galactocentric coordinate using Cartesian coordinates \\
 Z\_error&kpc & Uncertainty on z galactocentric coordinate using Cartesian coordinates \\
 SNe\_ratio& & Ratio of enrichment in the cluster from Type Ia vs. CCSNe supernovae \\
 SNe\_ratio\_erro& & Uncertainty on the ratio of enrichment in the cluster from Type Ia vs. CCSNe supernovae \\
 log\_age & dex & Log age in years of cluster, from \citet{cg2020}\\
 \hline
\end{tabular}
\end{table*}

\begin{figure*}
  \centering
  \includegraphics[width=\textwidth]{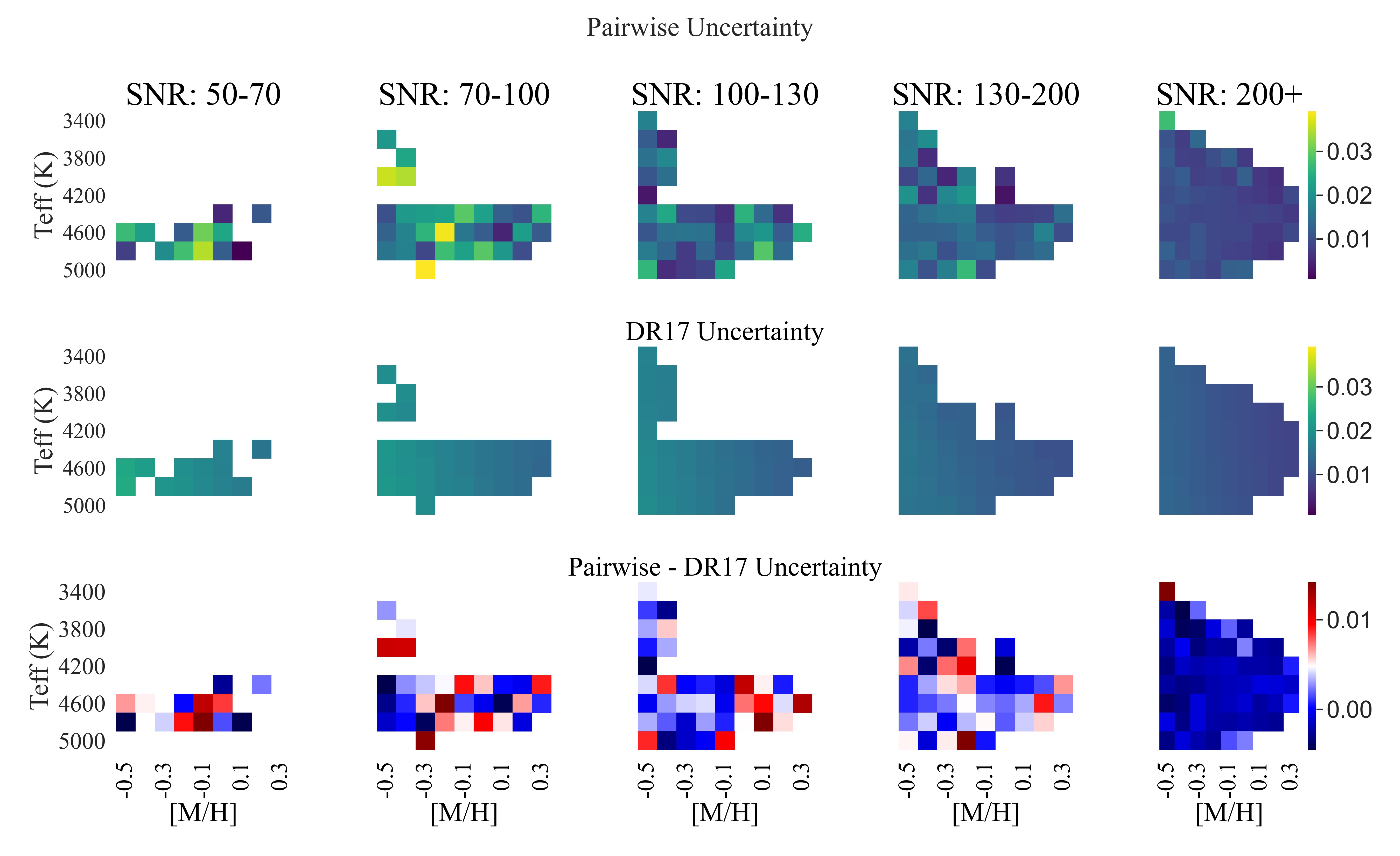}
  \caption{\emph{Top}: [Mg/Fe] uncertainties calculated using stars with multiple visits. \emph{Middle}: [Mg/Fe] uncertainties calculated by ASPCAP for DR17. \emph{Bottom}: The subtracted difference between the uncertainties calculated using multiple visits and the uncertainties from ASPCAP.}
  \label{fig:pairwise}
\end{figure*}
\subsection{Verification of Abundances $\&$ Uncertainties}
\label{sec:uncs}
The values of intrinsic scatter in OC abundances are on the order of $\sim$0.01~dex \citep[e.g.,][]{bovy,liu,ness,poovelil2020}, as are the abundance uncertainties within DR17. Therefore, verifying that the uncertainties on the abundances we are studying were not underestimated or overestimated was necessary. The method we use is the same as \citet{poovelil2020}. Since each star has some intrinsic true abundance measurement, repeated observations of the same star should result in a normal distribution around that value, where the width of the distribution is due only to the measurement uncertainties. Therefore we can use stars that have multiple visits in different APOGEE fields to measure the true uncertainty, and compare it to reported values within DR17 and IPL-3. To quantify our multiple-visit empirical uncertainty, we use Equation \ref{eqn:formula} from \citet{poovelil2020}.
\begin{equation}
\label{eqn:formula}
  e_{[X/Fe],k} = \frac{\sqrt{\pi}}{2}\rm{median}(|[X/Fe]_{j}-[X/Fe]_{i}]|),
\end{equation}

where $[X/Fe]_{j}$ and $[X/Fe]_{i}$ are the abundance measurements from the same star's \emph{i}th and \emph{j}th visits. $e_{[X/Fe],k}$ is the resulting [X/Fe] uncertainty after median stacking the pairwise measurements in the \emph{k}th bin. 

We group stars by SNR in bins spanning 50–70, 70–100, 100–130, 130–200, and greater than 200, as shown in Fig \ref{fig:pairwise}. Within each SNR range, stars are binned by $T_{\rm eff}$ and [M/H], where $\rm \Delta[M/H] = 0.2$~dex and $\Delta$ T$_{\rm eff}$ = 200 K. We use the K-S statistical test to ensure the distribution of abundance differences in each bin is consistent with a normal distribution, and flag those that are not to ensure they do not contaminate our sample. Bins that are not consistent with a normal distribution have two main causes. Firstly, some are at the edge of the [M/H] parameter space, in particular at low metallicities where ASPCAP is less reliable. Secondly, some bins have poor completion, with less than 10 measurements.

Within those bins where the empirical uncertainties are well-measured and normally distributed, we find very good agreement with the native pipeline uncertainties; thus for the rest of this work, we adopt those pipeline uncertainties directly from DR17 and IPL-3. 

\section{Measuring Chemical Scatter}
\label{sec:Meth}
\subsection{Paired Stars Method}
\label{sec:pairs}
Our primary method of determining the intrinsic scatter uses the difference in abundances between stars close to one another on the HR diagram: $\rm \Delta T_{\rm eff} < 100$~K and $\rm \Delta \log{g} < 0.1$~dex. These limits are chosen because the maximum induced abundance offset between pair members due to the systematics discussed in Section~\ref{sec:systematics} is on the order $\sim$0.001 dex.

We measure the intrinsic dispersion within each pair using Equation~\ref{eqn:pairwise_scatter}, where $e_{1}$ and $e_{2}$ are the abundance uncertainties of each star in the pair, $|\Delta[X/Fe]|$ is the absolute value of the difference in abundance measurements between the pair, and $\sigma$ is the inferred intrinsic dispersion.

\begin{equation}
\label{eqn:pairwise_scatter}
  \sigma = \sqrt{\frac{\frac{\pi}{2}|\Delta[X/Fe]|^2 - (e_{1}^2 + e_{2}^2))}{2}}
\end{equation}

We use a Monte Carlo method with N = 100 iterations to vary the abundance measurement of each star within the pair to estimate a final uncertainty on the measured pairwise scatter. Within each cluster, we separate the red clump and RGB stars as to not induce any scatter from potential evolutionary effects. Within each sub-sample, we sort the stars into pairs and then measure the intrinsic scatter between them. Finally, we take the median scatter of all the pairs within an OC as the true intrinsic scatter of the cluster. This method allows for extremely precise results, with final uncertainties on the order of $\sim$ 0.001 in most elements. 

To determine the number of pairs needed for a reliable measurement with this method, we use a synthetic “cluster” of points, with “true” and “observed” [X/H] abundances, and the same temperature and log(g) distribution as our real clusters. The true abundances for the synthetic stars reflect a given intrinsic scatter for the synthetic cluster. The observed abundances are generated by perturbing each true abundance by a random value drawn from a normal distribution with $\sigma$ set to the uncertainty of a real APOGEE star with the same temperature, log(g), and metallicity.

We then pair the stars as described in Section 3.1 and perform the intrinsic scatter measurement described using $N=3$ to $N=20$ pairs. We find that the measured cluster scatters are noisy and have both systematic offsets and larger uncertainties up until $N=8$, at which point the difference between the true and measured scatter does not change at larger $N$. Thus, we require a minimum of eight stellar pairs for clusters using this method.

Due to the restrictions outlined above regarding the separation of the pairs in $T_{\rm eff}$ and $\log{g}$, as well as the minimum number of pairs required, this method can only be applied to 15 of the 26 clusters studied in this paper. However due to the significantly higher precision of these values as compared to those derived using the method outlined in Section \ref{sec:mle_scatter}, within these 15 clusters we only publish results from this method.

\subsection{Maximum Likelihood Estimator}
\label{sec:mle_scatter}
\begin{figure}
  \centering
  \includegraphics[width=0.45\textwidth]{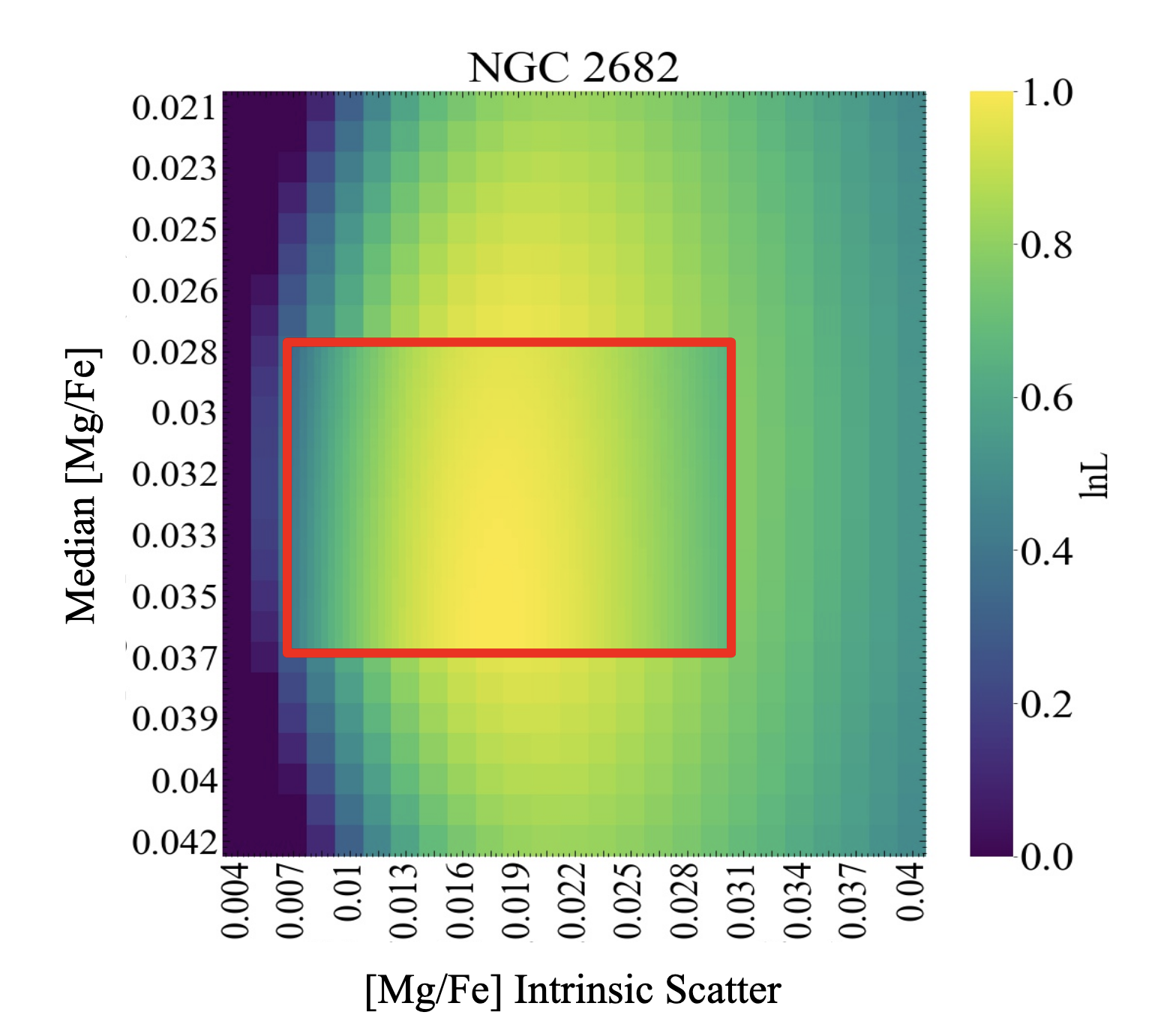}
  \caption{\emph{Outside Red:} The initial likelihood surface over which we initially test to determine a best guess for $\sigma_{[X/Fe]}$. \emph{Within Red:} Using the findings from the initial run, we more finely sample the likelihood surface in order to measure the final intrinsic scatter and uncertainty.}
  \label{fig:emle}
\end{figure}

We adopt a second method in the form of a Maximum Likelihood Estimator (MLE) to calculate the intrinsic scatter across our sample. This method produces larger uncertainties than our pairwise method, but it also has fewer sampling restrictions and can be applied to a larger set of OCs. The form of the MLE is shown in Equation \ref{eqn:mle} below.
\begin{equation}
\label{eqn:mle}
  \rm{ln}L = \prod_{i=1}^{\infty} \frac{1}{\sqrt{2\pi}(\sigma _{[X/Fe]}^{2} + e_{i}^{2})^{1/2}}exp(\frac{-(x_{i}-\mu _{[X/Fe]})^{2}}{2(\sigma _{[X/Fe]}^{2} + e_{i}^{2})}).
\end{equation}
In Equation \ref{eqn:mle}, $\sigma_{\rm [X/Fe]}$ is the intrinsic scatter being tested, and $e_i$ is the uncertainty on the [X/Fe] measurement for the $i$th star in that cluster. We sample a narrow range of mean [X/Fe] ($\mu$), where $\Delta\mu$ = 0.05 dex around the calculated mean of the cluster members, with an initial range of 0.1~dex for intrinsic scatter and spacing of 0.003~dex. We then do a second iteration with finer spacing in the intrinsic scatter dimension, with spacing on the order of $10^{-4}$~dex, centered on the likeliest value from the coarser grid. An example of this is shown in Figure \ref{fig:emle}. We calculate a variance from the Fisher information matrix, and from that we calculate the uncertainty the intrinsic scatter. We apply the MLE method to all twenty-six clusters in our sample. Of these twenty-six clusters in our sample, ten
use the MLE intrinsic scatters for their final measurement.

\subsubsection{MLE Corrections for Small Samples}
\label{sec:corrections}
Based on tests with synthetic data, we find that the measured intrinsic scatter is unreliable when the number of cluster stars ($N$) is low, with a consistent systematic bias at $N<9$ in the measured scatter. This is due to the fact that the MLE is a biased estimator and at small sample sizes has a negative bias that causes it to underestimate the true parameter value, which can be accounted for with a multiplicative factor \citep{bias_correction}: 

\begin{equation}
\label{eqn:downturn}
  \Delta\sigma_{true} = \sqrt{\frac{N}{N-1}}\sigma_{MLE}
\end{equation}

This correction is applied to clusters with 6 $\leq$ N $<$ 9 members. Below six members, the measured scatters are too unpredictably noisy. This small sample size correction results in an additional six clusters added to our sample, creating the final sample of 26 OCs.

\subsubsection{Systematics}
\label{sec:systematics}
As discussed in \citet{ASPCAP2016}, the methodology for measuring a star's [X/Fe] involves a multi-parameter fit that includes fitting the observed stellar spectrum to synthetic models. First a global fit to the spectrum is done to determine the best fit values for temperature, surface gravity, microturbulent velocity, and [M/H]. Holding these parameters constant, individual elemental abundances are extracted from narrow spectral windows. 

\begin{figure}[hbt!]
  \centering
  \includegraphics[width=0.45\textwidth]{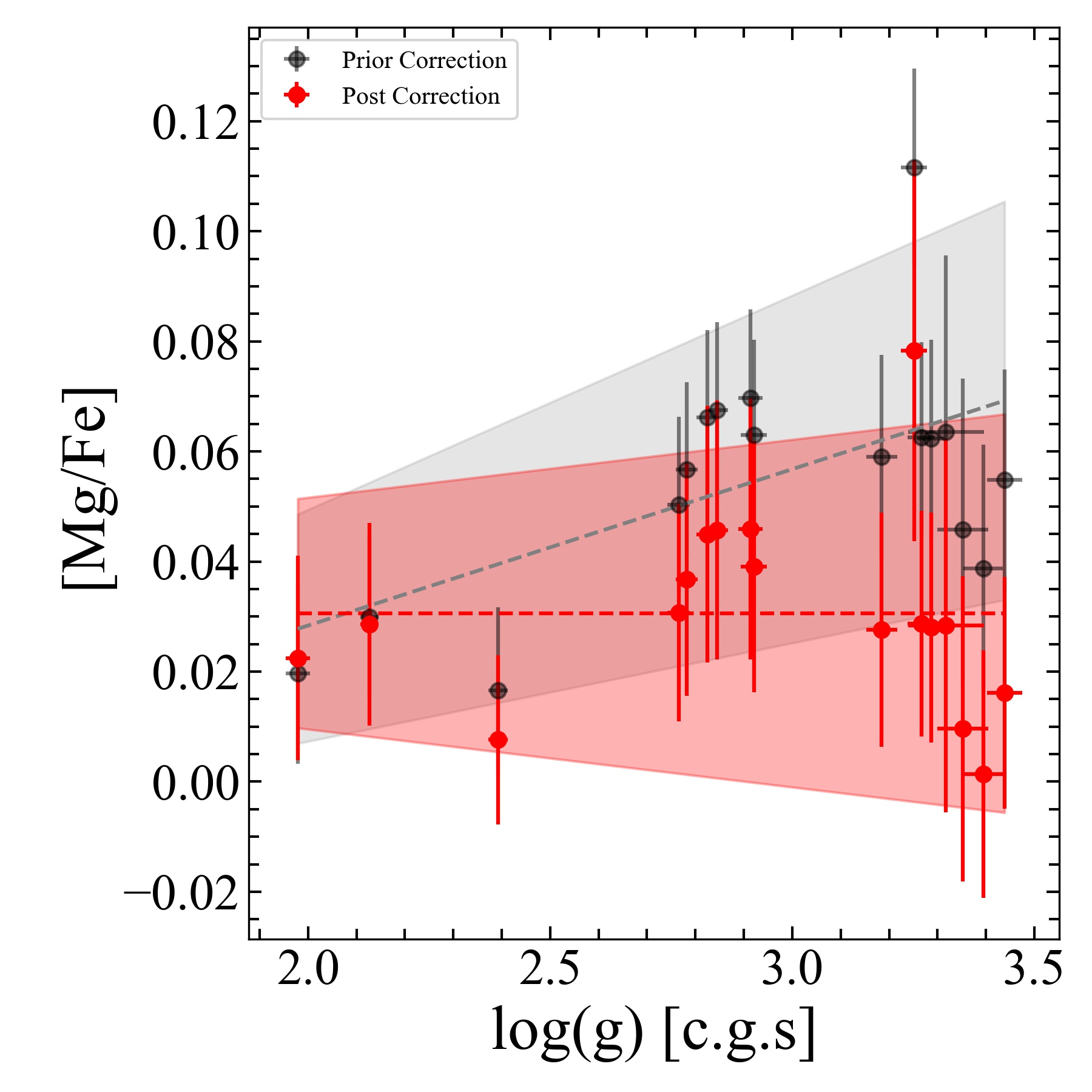}
  \caption{Uncalibrated surface gravity versus [Mg/Fe] for M67. The [Mg/Fe] values from DR17 are shown in black, and have a clear linear relationship with log(g). The adjusted abundances used in the calculation of intrinsic scatter are shown in red.
  }
  \label{fig:stellar_params_corr}
\end{figure}

To test for any systematic trend between the global stellar parameters and elemental abundances, we quantify the slope of the uncalibrated log(g) vs. [X/Fe] within all the clusters in each pipeline, approximating it as a linear relationship, as seen in Figure \ref{fig:stellar_params_corr}. We exclude red clump stars as they are further along their evolutionary track than red giant stars, and potentially have slightly different surface abundances due to evolutionary effects or internal systematics. As a result, including them in the slope measurement could artificially drive any measurements of chemical homogeneity. However, these RC stars are adjusted afterwards along with the other giants in their cluster. This process ensures a uniform abundance correction across the entire cluster. From this, we find that the existing slopes in the cluster sample are nonzero, with a median slope of $\sim$0.02 dex/dex. 

This systematic bias is present in DR17, and in the \emph{Cannon} and ASPCAP allStar files from IPL-3. However it is not present in the IPL-3 allStar release analyzed using the \emph{Payne}. We adjust the measured [X/Fe] for each cluster star using the following equation:

\begin{equation}
  [X/Fe]_{i,\ corrected} = [X/Fe]_{i} - m\rm{\log{g}_{\emph{i}}} + ZP,
\end{equation}

where each \emph{i}th index is a star in a cluster and $m$ is the best fit slope of [X/Fe] and log(g) within a specific cluster. We set the zero-point (ZP) of the cluster [X/Fe] after the correction using the abundances of the stars on the giant branch below the 
red clump to ensure that the median cluster [X/Fe] is reflective of its true value. The fitting uncertainties are propagated to uncertainties on the correction, which are then added in quadrature with the existing abundance uncertainties. 

\section{Results}
\label{sec:Res}
Within one standard deviation, the only abundance that showed evidence of inhomogeneity, or consistent nonzero intrinsic scatter, across multiple clusters was [M/H]. This is because [M/H] has small uncertainties compared to other [X/Fe] measurements ($\sim$0.008 dex as compared to $\sim$0.015 dex for the rest of the DR17 and IPL-3 abundances and $\geq$0.03 for the BAWLAS abundances). Given that the [M/H] uncertainties are smaller but not under-reported, as we verified in Section 2, it potentially implies that we could detect the existence of inhomogeneities that were then masked by larger uncertainties in the other abundances. 

However, within three standard deviations (a 99.7$\%$ confidence interval), none of the clusters show measurable inhomogeneities in any of the measured elements. Furthermore, the scale of the limits derived using the paired stars method are also on the scale of $\sim$0.001 dex in many elements. And using that method we find no measurement of scatter across in any of the elements and clusters. As a result, we are confident that in the majority of elements we can constrain the homogeneity of the OCs to less than 0.01~dex in a 99.7$\%$ confidence interval.

\begin{figure*}[h!]
  \includegraphics[width=\textwidth]{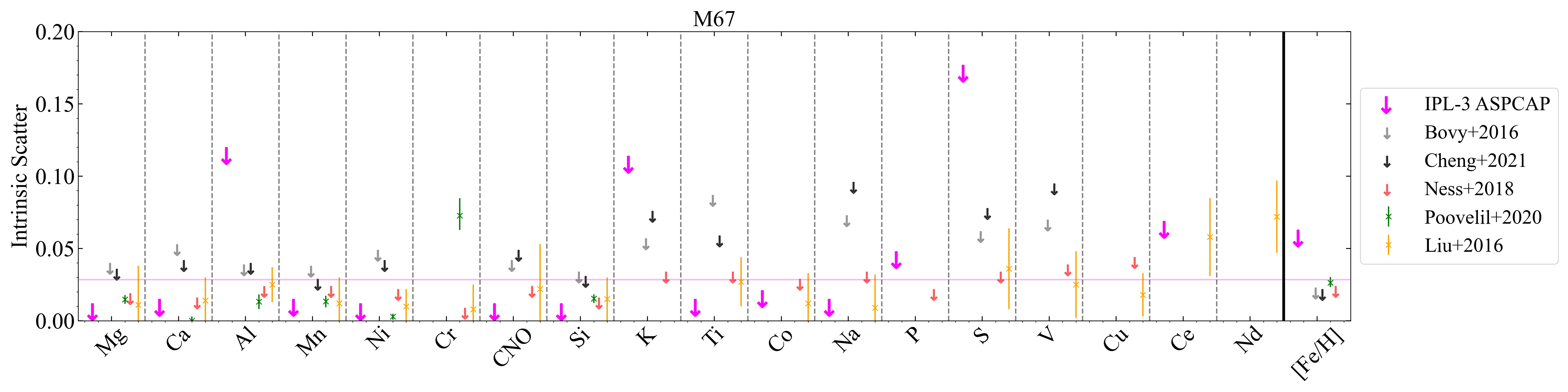}
  \includegraphics[width=\textwidth]{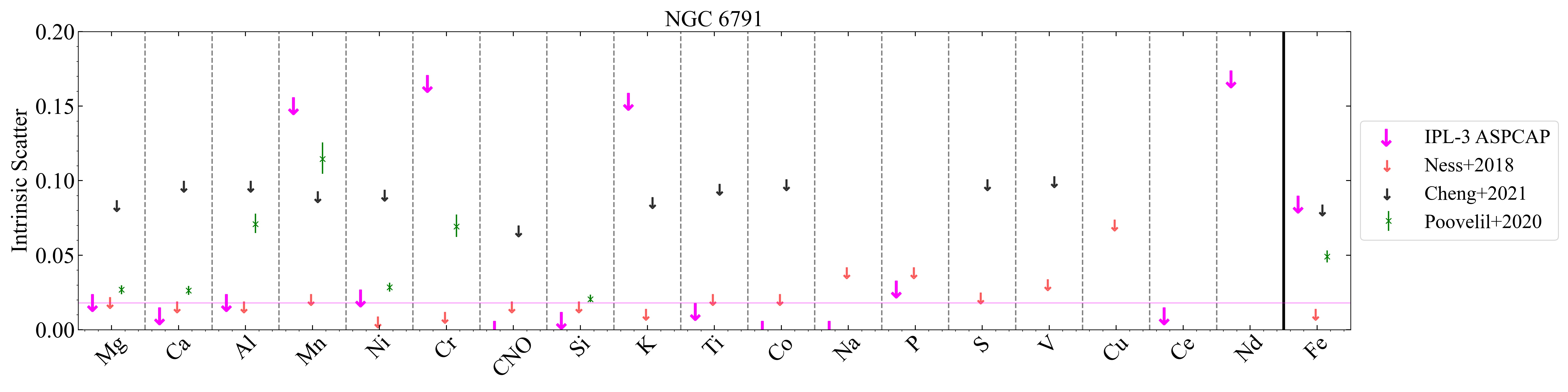}
  \includegraphics[width=\textwidth]{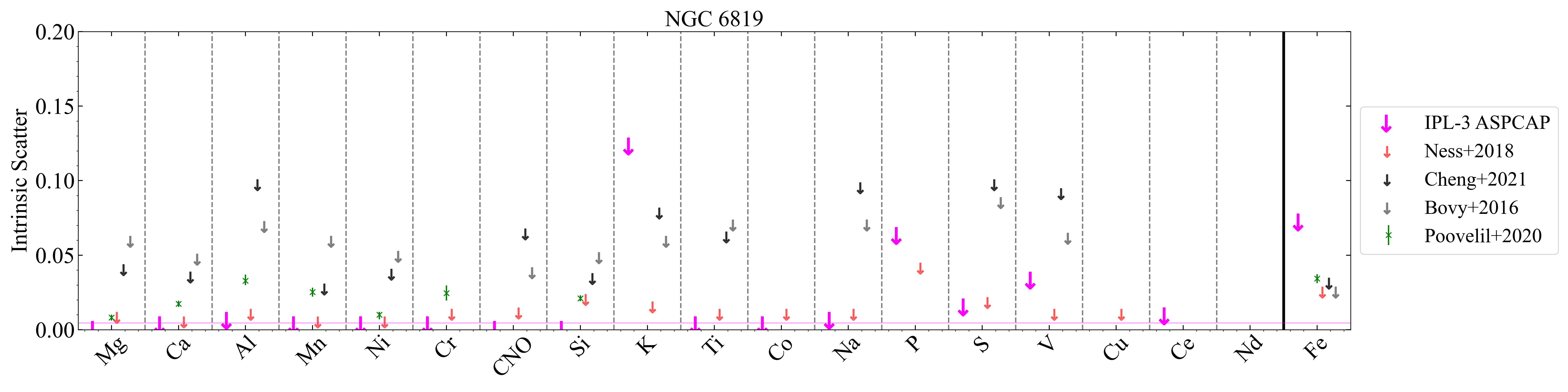}
  \caption{The 99.7\% upper limits using abundances from IPL-3 ASPCAP in M67, NGC~6791, and NGC~6819. The median 99.7~percentile upper limit for each cluster is shown as a magenta horizontal line. We compare our findings to upper limits and measurements from previous literature. The upper limits are broadly in agreement with other studies for well-measured elements in APOGEE and Milky Way Mapper.}
  \label{fig:m67}
\end{figure*}

All the measured quantities for each [X/Fe] are presented in Table~\ref{table2}. We show that across DR17 and all MWM pipelines, we do not find any elements that show consistent chemical inhomogeneity across our cluster sample, and in Figure \ref{fig:m67} we show that we do not find any clusters with consistent scatter across their abundance samples. We only show the results from the IPL-3 ASPCAP data in the here because it includes values for weak lined elements. The literature comparison plots using DR17, IPL-3 \emph{Cannon}, and IPL-3 \emph{Payne} releases are comparable and are shown in Appendix~\ref{sec:appendix_a}. 

\begin{table*}
\centering
\caption{List of [X/Fe] columns in the table of cluster parameters for IPL-3 ASPCAP. All columns have units of dex, and all X/Fe measurements are scaled to solar values. The abundances measured in these columns are Mg, Ca, Si, S, P, Ni, Cr, Ti, V, Mn, Co, Na, K, Al, V, Mn, Co, Cu, CNO, Nd, and Ce. In addition we provide columns for [Fe/H] and [M/H]. This table will be made available in machine readable format.}
\begin{tabular}{l|l}
\label{table2}
 Column Name& Column Description \\
 \hline
 \hline
 cluster & Cluster name\\
 Pipeline & Which pipeline (ASPCAP, \emph{Cannon}, \emph{Payne}) was used to derive abundances \\
 Method & Whether the paired stars (\texttt{pair}) or MLE method (\texttt{MLE}) was used\\
 x\_fe & Median [X/Fe] of the cluster \\
 x\_fe\_cluster\_disp & Dispersion in [X/Fe] within the cluster\\
 x\_fe\_sigma & [X/Fe] intrinsic MLE scatter \\
 x\_fe\_sigma\_uncertainty & [X/Fe] intrinsic pairwise scatter uncertainty \\
 x\_fe\_sigma\_upper\_limit & [X/Fe] intrinsic MLE scatter 3 uncertainty upper limit \\
 \hline
\end{tabular}
\end{table*}

The upper limits on the intrinsic scatter measured in elements included in the BAWLAS catalog are higher than the limits on intrinsic scatter placed on the more well-measured elements, such as Mg or Ni. The reasons for this are twofold: Firstly, while many of the clusters studied did have enough stars to measure an intrinsic scatter, the number of stars that contained BAWLAS abundances within each cluster was smaller than the number of stars used to calculate $\alpha$ and iron-peak elements. Secondly, the associated uncertainties for the weak-lined elements were appreciably larger (0.03--0.08 dex) than the ones included in DR17 and MWM (0.01--0.04 dex). 

\section{Discussion}
\label{sec:Disc}
\subsection{Comparison to Milky Way Field Stars}
\label{sec:mfs}
\begin{figure*}
  \includegraphics[width=\textwidth]{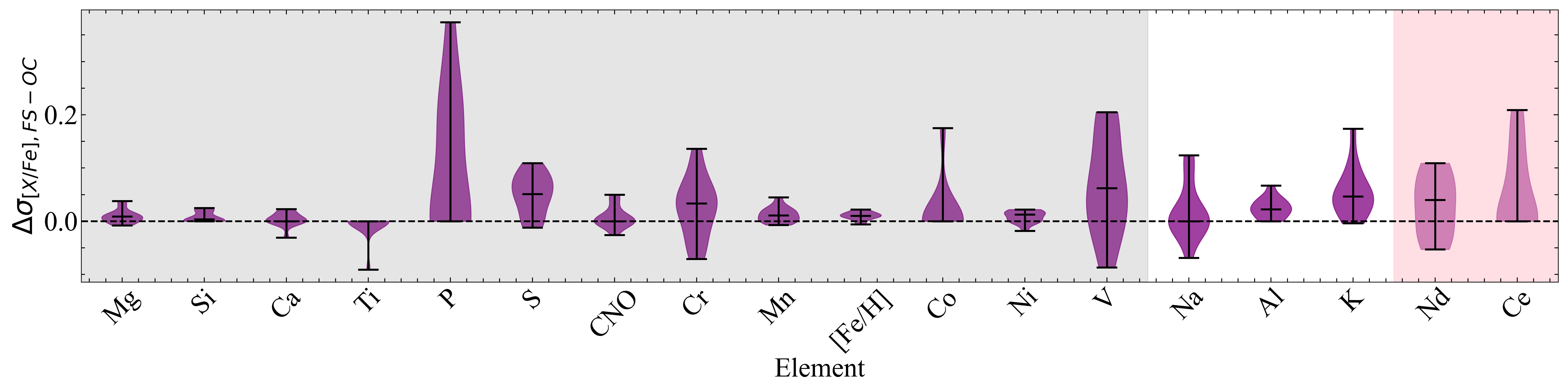}
  \caption{A violin plot comparing the difference in element intrinsic scatter between open clusters and matched samples of field stars as shown in Section \ref{sec:mfs}. Here the elements that belong to the $\alpha$, CNO, and iron-peak families are shaded in grey as we expect them to be similar due to our selection criteria. Within these elements P and V have larger limits due to them being weak lined elements from the BAWLAS catalog. While broadly we find that the field star samples have more scatter than their OC counterpart, we don't find any conclusive difference in any element between the two samples.}
  \label{fig:field_star_comparison}
\end{figure*}
To quantify the difference in chemical homogeneity between our OCs and the Milky Way field, we create a matched field star sample (MFS) that mirrors our existing cluster sample.

For each of the 15 clusters in our study with enough members to apply the pairwise method, we match each star within the cluster to a field star within two sigma\footnote{\bf We tested the effects of using 1$\sigma$ and 3$\sigma$ to match stars and found no difference in our conclusions.} in Galactocentric radius, [M/H], [$\alpha$/M], T$_{\rm eff}$, and $\log{g}$. Here we consider the parameter uncertainties of both the cluster star as well as any candidate field stars. We use each star's [M/H] and [$\alpha$/M] value from prior to the stellar parameter correction. We then measure the intrinsic scatter in each of the MFS samples, replicating the methodology outlined in Section~\ref{sec:pairs}. Finally, we compare the difference in intrinsic scatter between our MFS sample and the OC sample (Figure~\ref{fig:field_star_comparison}). 

We find that on average, across all abundances, the matched field star samples have $+$0.012$_{-0.01}^{+0.02}$~dex more intrinsic scatter than the open clusters in our sample. This is in strong agreement with \citet{ness2021}, which states that stars in the Milky Way are chemically similar (within $\sim$0.01--0.02 dex) when given a fixed Galactocentric radius, [M/H], and [$\alpha$/Fe]. The median difference between OC intrinsic scatter and field star intrinsic scatter ($\Delta\sigma_{[X/Fe]}$) for each nucleosynthetic family is given below:

\begin{itemize} \itemsep -2pt
  \item $\alpha$-elements (Mg, Si, Ca, Ti, P, S): 0.002 dex
  \item Iron-peak elements (Cr, Mn, Fe, Co, Ni, V): 0.012 dex
  \item Odd-Z (Na, Al, K): 0.023 dex
  \item Neutron-capture (Nd, Ce): 0.02 dex
\end{itemize}

Due to our selection criteria for the field star sample, we expect similar intrinsic scatter in the $\alpha$, and iron-peak elements. 
Interestingly, two of the odd-Z elements, Al and K, both have nonzero scatter in the majority of our MFS samples despite being measured as homogeneous in our OCs. While in [Na/Fe] the distribution in $\Delta\sigma_{[X/Fe]}$ is roughly symmetric, Na is an element with weak lines in APOGEE's wavelength range and was included in our sample with the BAWLAS catalog. As a result, not only does it have comparatively larger uncertainties than the other odd-Z elements, but it has poorer completion as well, as only a subset of high SNR stars in our study have BAWLAS abundances. This implies that odd-Z elements may be a useful tool in differentiating otherwise chemically similar populations. However, given that the distribution for all of these elements is consistent with zero in at least a subset of clusters and field star comparisons, more precise limits are needed to accurately test this 

Neutron capture elements also show slightly larger differences in scatter than their field star counterparts. \citet{manea2023} showed that neutron-capture elements have more discriminatory power in distinguishing ``doppelganger stars''. What we find potentially corroborates that, but we also show that for co-eval and co-natal stars within an OC, the expectation of chemical homogeneity within neutron capture elements is broadly comparable to that of elements from other well-studied and well-measured nucleosynthetic families. In a field star sample with a high expectation of chemical similarity to an OC, as shown by the relative lack of difference in measured $\alpha$-element intrinsic scatter, there is a $0.02 \pm 0.02$~dex difference in measured scatter for neutron capture elements. Due to this, it seems possible that neutron capture elements could be useful in distinguishing otherwise highly similar stellar populations but significantly more precise limits would be required to accomplish that goal.

In both cases of odd-Z and neutron capture elements, the differences between each OC and their respective MFS are consistent with zero within 3 uncertainties. More precise limits are required to make any conclusive statements on their effectiveness in distinguishing co-natal and co-eval stellar populations.

\subsection{Previous Studies of Chemical Homogeneity}
There have been numerous studies focusing on measuring the chemical homogeneity of star clusters; however, most of these studies have been focused on larger and more complicated globular clusters. Within open clusters, different studies have found a wide range of limits on inhomogeneity. This is further complicated by the fact that not every study investigates the same set of abundances, nor is every analysis method comparable to one another. One of the differences between this study and many others is that their published limits on homogeneity are either 68$\%$ and 95$\%$ limits; the values we publish and show in Figures \ref{fig:m67} and Appendix \ref{sec:appendix_a} are all 99.7$\%$ limits.

Given the number of studies done on M67, NGC 6791, and NGC 6819, they are shown in Figure \ref{fig:m67}, while the literature comparisons for the remaining clusters are located in Appendix \ref{sec:appendix_a}. The figures comparing the intrinsic scatter measured using DR17, \emph{Cannon}, and \emph{Payne} abundances for all 26 OCs are also shown in Appendix \ref{sec:appendix_a}.

We compare our results to four previous studies \citep{bovy,liu,ness,poovelil2020}. We find that within the $\alpha$ and iron peak elements, which are well-measured in APOGEE and MWM, the upper limits derived in this study are consistent with previous findings. In the weak-lined elements such as V, Cu, Ce, and Nd, there is more variance. But even within those elements, in many clusters we find comparable limits to previous works.

It is worth outlining the differences in the analyses and sample sets between these different studies. \citet{poovelil2020} is the most similar to ours in terms of sample size, analysis method, and dataset \citep[using APOGEE DR16;][]{dr16}. While they measure upper limits that are far less constraining than ours, they also measure {\it lower} limits that strongly imply the existence of true intrinsic scatter. However, the stellar parameter systematic that we found in DR17 and MWM is also present in DR16 but was unaccounted for in \citet{poovelil2020}'s final results. Therefore, it is possible that the measured scatters in \citet{poovelil2020} are impacted by a relationship between stellar parameters and abundance measurements. Our uncorrected $\sigma_{[X/Fe]}$ values (not shown in this paper) are in strong agreement with the ones published in \citet{poovelil2020}, which lends evidence to this hypothesis.

\citet{liu} uses high resolution spectroscopy ($R \sim 50,000$) to study two solar twins in M67. This, makes it less likely that their final abundances are driven by systematics due to stellar atmospheres or poor line fitting, as the stars are in very similar parts of parameter space. This method is similar, but not identical, to the pairwise method of deriving intrinsic scatter outlined in Section \ref{sec:pairs}. They derive abundances for a total of 26 elements as well as [Fe/H], with an average measurement uncertainty of e$_{[X/Fe]}$ $\leq$ 0.02 dex. As a result, within their sample they were able to very tightly constrain the difference in [X/Fe] between the two stars, showing that for elements with $Z<30$, there was an average difference of $<$ 0.06 dex, and for the neutron capture elements there was an average difference of $<$ 0.05 dex. It is encouraging that in the strong-lined elements included in our study, we place similar upper limits on abundance scatter.

\citet{ness} uses APOGEE DR13 \citep{dr13} abundances and spectra to measure intrinsic dispersions within a set of seven open clusters, six of which are included in this study. However, their methodology in constraining scatter was noticeably different than this work. Using \emph{The Cannon} \citep{cannon}, they derive abundances and uncertainties in 20 different elements from APOGEE spectra and a training set of open cluster stars in APOGEE DR13. \citet{ness} notes that while their \emph{Cannon} abundance measurements were broadly comparable to ASPCAP's, the uncertainties are between 20-50\% smaller. Beyond that, using a chi-squared fit they determined that the uncertainties, a quadrature sum of formal abundance uncertainty and cross-validation uncertainty, are overestimated given the widths of calculated abundance distributions. Therefore, they derive a scaling factor to correct the uncertainties to match the theoretically predicated value; however, this methodology also introduces the risk of artificially down-scaling the measured limits on the intrinsic scatter. However, it is also worth noting that in the majority of elements, the values they publish are comparable to the ones derived in this work.

\citet{bovy} studied the abundance spread of 15 different elements in three clusters, all of which are included in this study: M67, NGC~2420, and NGC~6819. The data came from APOGEE DR12 \citep{dr12}, with cluster membership from \citet{mezasros2013}. However, the key difference between their study and this one is the methodology. \citet{bovy} made the assumption that in the absence of any intrinsic chemical scatter, the main driver for variation in the photometric and spectroscopic attributes of OC stars is their mass, which can be modeled as a one dimensional sequence. They correct for any systematic variations in the spectra driven by mass, using $T_{\rm eff}$ as a proxy. They then use detailed forward modeling of the spectra and Approximate Bayesian Computation to measure the intrinsic scatter as well as upper limits. Overall \citet{bovy} find no indication of chemical inhomogeneity in any of the three clusters they studied; the upper limits they derived are largely in agreement with the ones calculated using the MLE method in this work. However, using the paired stars method we can constrain tighter upper limits in the majority of elements. Similar to studies discussed previously, the limits placed on the BAWLAS neutron capture elements by \citet{bovy} are lower than ours.

Finally, \citet{cheng2021} use spectroscopic data from APOGEE DR14 \citep{dr14} to measure intrinsic scatter in M67, NGC 6791, and NGC 6819 in 15 different elemental abundances. The analysis method is very similar to the one outlined in \citet{bovy}, though there are a few differences --- notably, that they use DR14 instead of DR12, which includes several differences in the line lists \citep[detailed in][]{dr14_analysis}. Furthermore, unlike \citet{bovy}, they use spectroscopic effective temperatures in their one-dimensional model as opposed to photometric effective temperatures. \citet{cheng2021} found the clusters to be chemically homogeneous, placing upper limits comparable to this study across its sample of elements. While they measured fewer abundances than this work, in the three clusters studied the upper limits they derive are similar to \citet{bovy}.

Thus, within the majority of the elements included in DR17 and IPL-3, such as the $\alpha$-elements and iron-peak elements, the upper limits we calculate here are in agreement with what has been previously found across all the clusters studied. This lends credibility to the limits placed on the numerous clusters studied in this work that did not have previously derived limits in the literature. However, the limits derived in this study for the neutron capture elements are larger than what has been previously found in any of the literature. This is likely driven by the comparatively large uncertainties on the elements (0.03--0.08 dex).

\section{Conclusion}
\label{sec:conc}
The purpose of this study was to quantify the level of chemical homogeneity within the largest sample to date of Milky Way open clusters for a broad set of elements. Using SDSS-V Milky Way Mapper IPL-3 abundances and Gaia DR3 kinematics, we identify a sample of 26 open clusters with large enough membership to measure the intrinsic scatter in up to 20 elements. Using the abundance differences between paired stars along the HR diagram, as well as a Maximum Likelihood Estimator, we then measure the intrinsic scatter within each element for each cluster. We find the following:
\begin{enumerate}
 \item We assemble a sample of 26 open clusters across a broad range of metallicity, age, mass, and galactic radii. Within a 99.7\% confidence interval, we do not find any evidence of intrinsic scatter on the giant branch or in the red clump in any element across all the open clusters in our sample. 
 
 \item Within the majority of abundances included in APOGEE DR17 and Milky Way Mapper IPL-3, we constrain the chemical homogeneity to $\leq$0.02~dex within a 99.7$\%$ confidence interval, and within $\leq$0.2~dex for the weak lined elements, such as those included in the DR17 BAWLAS catalog. Our limits are consistent with those in the literature for well-studied elements and clusters, and we add roughly a dozen clusters to this literature sample. Given the limited dataset in some of the elements, we recommend follow up measurements to better quantify their upper limits.
 
 \item When compared to a sample of field stars with similar Galactocentric radii, [$\alpha$/M], and [M/H], we find our OCs to be more chemically homogeneous, with an average difference of $\sim$0.012 dex between the two samples. This corroborates previous findings that the dimensionality of chemical enrichment of the Milky Way is low, and can likely be explained through a few processes. In the future this could be useful in placing constraints on radial mixing and azimuthal variations within the Milky Way.

 \item We identify surface-gravity-dependent abundance shifts within APOGEE DR17 and Milky Way Mapper IPL-3 (corrected for in this analysis). This systematic needs to be accounted for in similar future work. We also find that the abundance uncertainties within both APOGEE and MWM are accurately estimated.

 \item These findings have implications for attempts to implement chemical tagging, especially strong chemical tagging, specifically showing that within the light elements alone it is not possible to confidently separate field stars and co-natal stars given similar stellar parameters and Galactic radii. The tightest abundance variation constraints in OCs may also help set limits on the rate of binary interactions and planetary engulfment in different environments.

\end{enumerate}

\section{Acknowledgments}
We thank the anonymous referee for their helpful and insightful comments. This material is based upon work supported by the National Science Foundation under Grant No. AST-2206542. A.S. also acknowledges support from the University of Utah's J. Irvin and Norma K. Swigart First-Year Summer Graduate Research Fellowship. P.F. acknowledges support from NSF Grant AST-2206541, and K.C. acknowledges support from NSF Grant AST-2206543.

Funding for the Sloan Digital Sky Survey IV has been provided by the Alfred P. Sloan Foundation, the U.S. Department of Energy Office of Science, and the Participating Institutions. 
SDSS-IV acknowledges support and resources from the Center for High Performance Computing at the University of Utah. The SDSS website is \url{www.sdss4.org}.
SDSS-IV is managed by the Astrophysical Research Consortium for the Participating Institutions of the SDSS Collaboration including the Brazilian Participation Group, the Carnegie Institution for Science, Carnegie Mellon University, Center for Astrophysics | Harvard \& Smithsonian, the Chilean Participation Group, the French Participation Group, Instituto de Astrof\'isica de Canarias, The Johns Hopkins University, Kavli Institute for the Physics and Mathematics of the Universe (IPMU) / University of Tokyo, the Korean Participation Group, Lawrence Berkeley National Laboratory, Leibniz Institut f\"ur Astrophysik Potsdam (AIP), Max-Planck-Institut f\"ur Astronomie (MPIA Heidelberg), Max-Planck-Institut f\"ur Astrophysik (MPA Garching), Max-Planck-Institut f\"ur Extraterrestrische Physik (MPE), National Astronomical Observatories of China, New Mexico State University, New York University, University of Notre Dame, Observat\'ario Nacional / MCTI, The Ohio State University, Pennsylvania State University, Shanghai Astronomical Observatory, United Kingdom Participation Group, Universidad Nacional Aut\'onoma de M\'exico, University of Arizona, University of Colorado Boulder, University of Oxford, University of Portsmouth, University of Utah, University of Virginia, University of Washington, University of Wisconsin, Vanderbilt University, and Yale University.

Funding for the Sloan Digital Sky Survey V has been provided by the Alfred P. Sloan Foundation, the Heising-Simons Foundation, the National Science Foundation, and the Participating Institutions. SDSS acknowledges support and resources from the Center for High-Performance Computing at the University of Utah. The SDSS web site is \url{www.sdss.org}. SDSS is managed by the Astrophysical Research Consortium for the Participating Institutions of the SDSS Collaboration, including the Carnegie Institution for Science, Chilean National Time Allocation Committee (CNTAC) ratified researchers, the Gotham Participation Group, Harvard University, Heidelberg University, The Johns Hopkins University, L’Ecole polytechnique fédérale de Lausanne (EPFL), Leibniz-Institut für Astrophysik Potsdam (AIP), Max-Planck-Institut für Astronomie (MPIA Heidelberg), Max-Planck-Institut für Extraterrestrische Physik (MPE), Nanjing University, National Astronomical Observatories of China (NAOC), New Mexico State University, The Ohio State University, Pennsylvania State University, Smithsonian Astrophysical Observatory, Space Telescope Science Institute (STScI), the Stellar Astrophysics Participation Group, Universidad Nacional Autónoma de México, University of Arizona, University of Colorado Boulder, University of Illinois at Urbana-Champaign, University of Toronto, University of Utah, University of Virginia, and Yale University.

\software{astropy \citep{astropy:2013,astropy:2018,astropy:2022}, scipy \citep{scipy2020},matplotlib \citep{matplotlib}, numpy \citep{numpy}, pandas \citep{pandas_1, pandas_2}, and seaborn \citep{seaborn}.}

\bibliography{references}
\appendix
\section{}
\label{sec:appendix_a}
Here we show the literature comparison plots for the well studied OCs in our sample, as discussed in Section \ref{sec:Disc}.

\begin{figure*}[hbt!]
  \centering
  \includegraphics[width=0.9\textwidth]{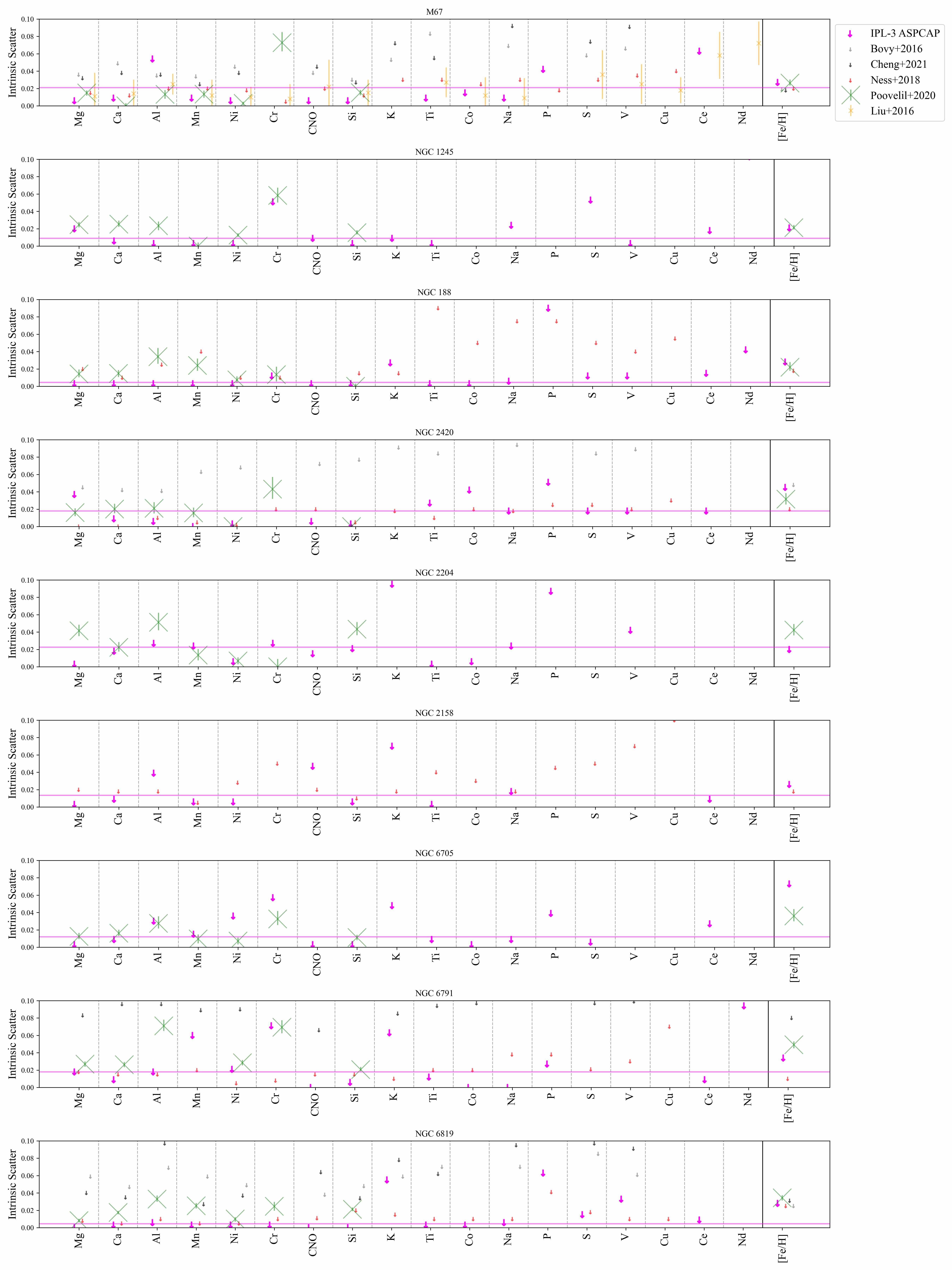}
  \caption{The literature comparison plots for the well studied OCs in our sample using IPL-3 ASPCAP (colored in magenta). The horizontal line indicates the median 99.7$\%$ confidence limit for that cluster's homogeneity.}
\end{figure*}
\begin{figure*}[hbt!]
  \centering
  \includegraphics[width=0.9\textwidth]{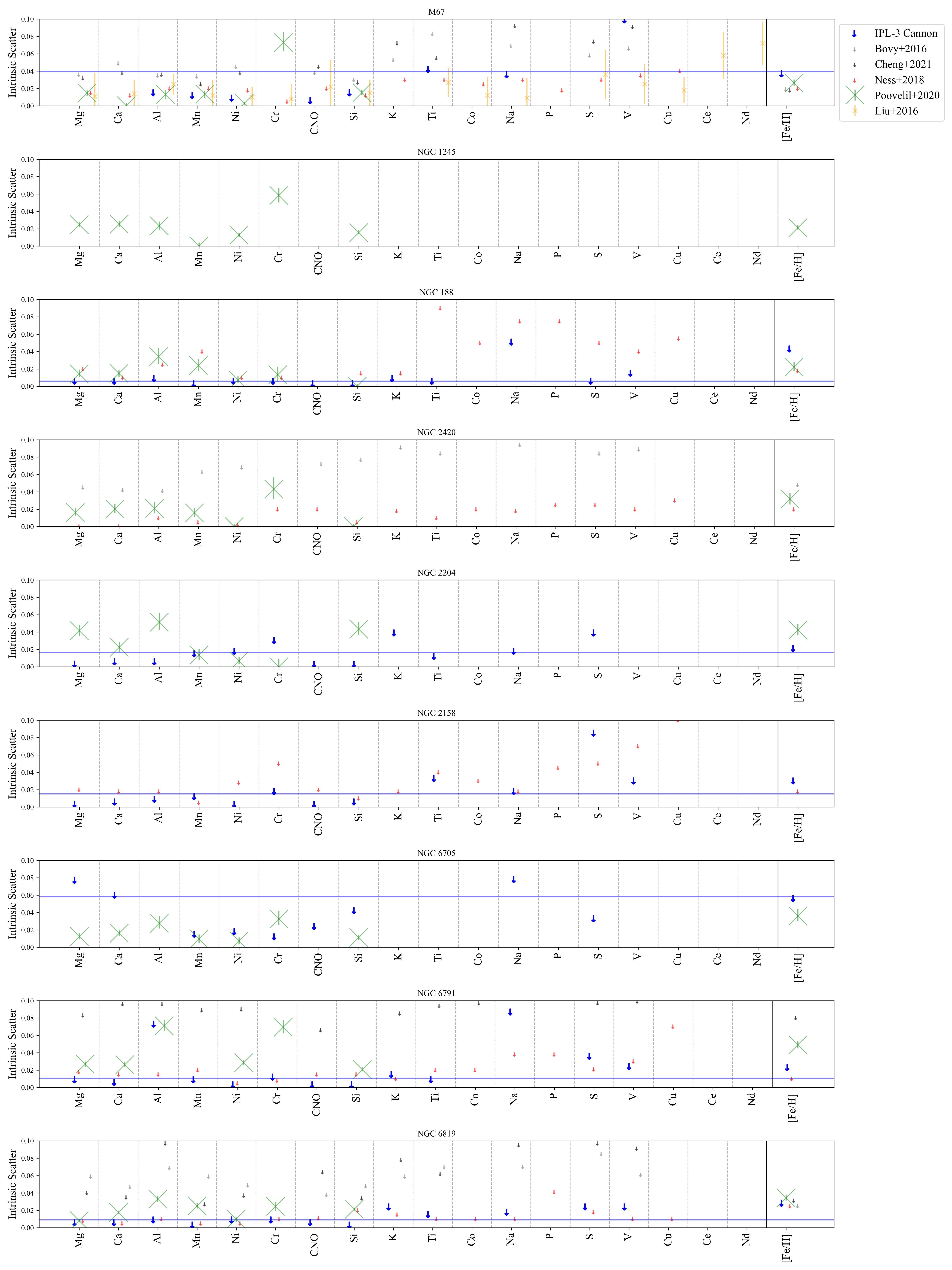}
  \caption{The literature comparison plots for the well studied OCs in our sample using IPL-3 \emph{Cannon} (colored in blue). The horizontal line indicates the median 99.7$\%$ confidence limit for that cluster's homogeneity.}
\end{figure*}
\begin{figure*}[hbt!]
  \centering
  \includegraphics[width=0.9\textwidth]{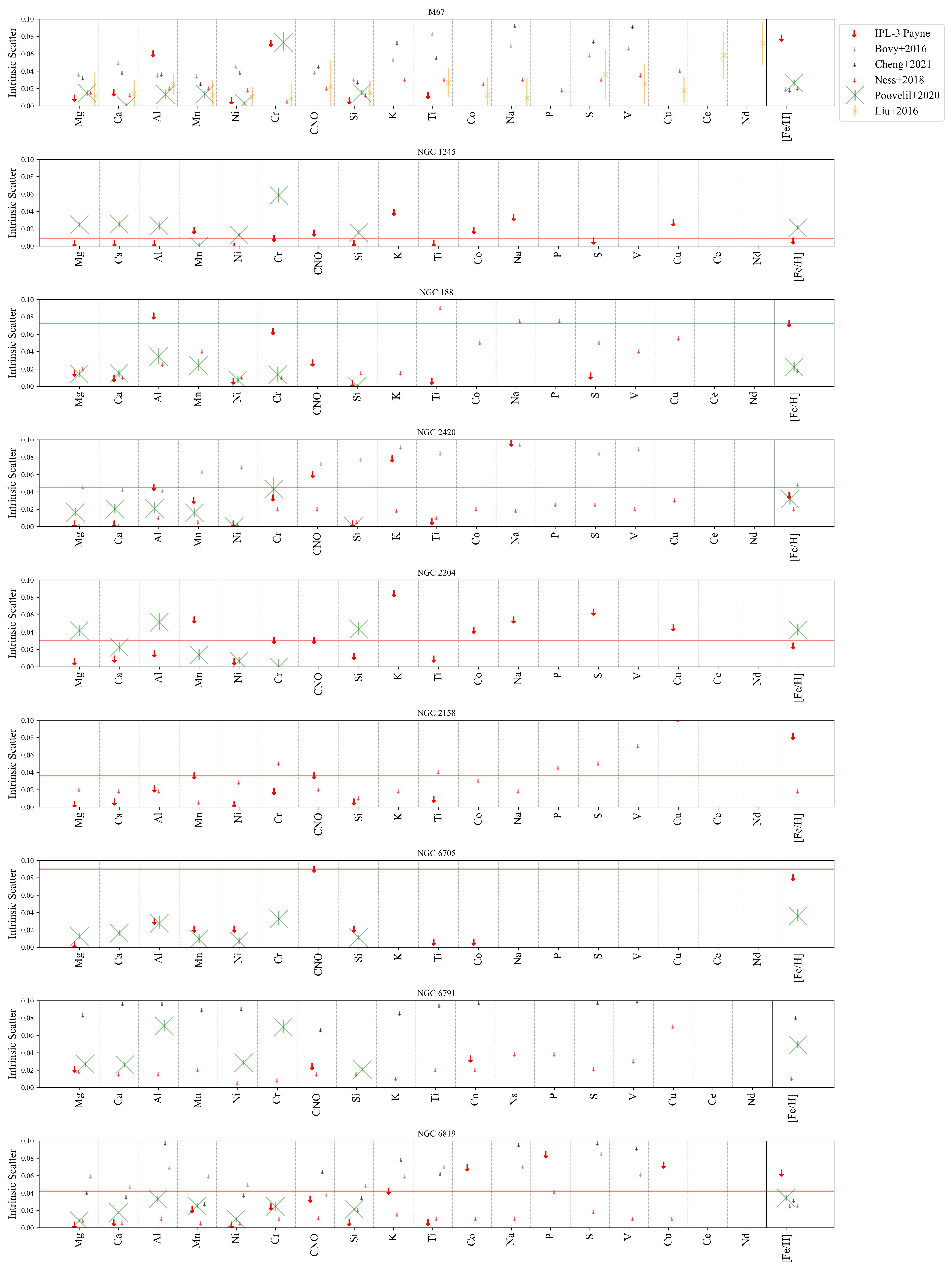}
  \caption{The literature comparison plots for the well studied OCs in our sample using IPL-3 \emph{Payne} (colored in red). The horizontal line indicates the median 99.7$\%$ confidence limit for that cluster's homogeneity.}
\end{figure*}
\begin{figure*}[hbt!]
  \centering
  \includegraphics[width=0.9\textwidth]{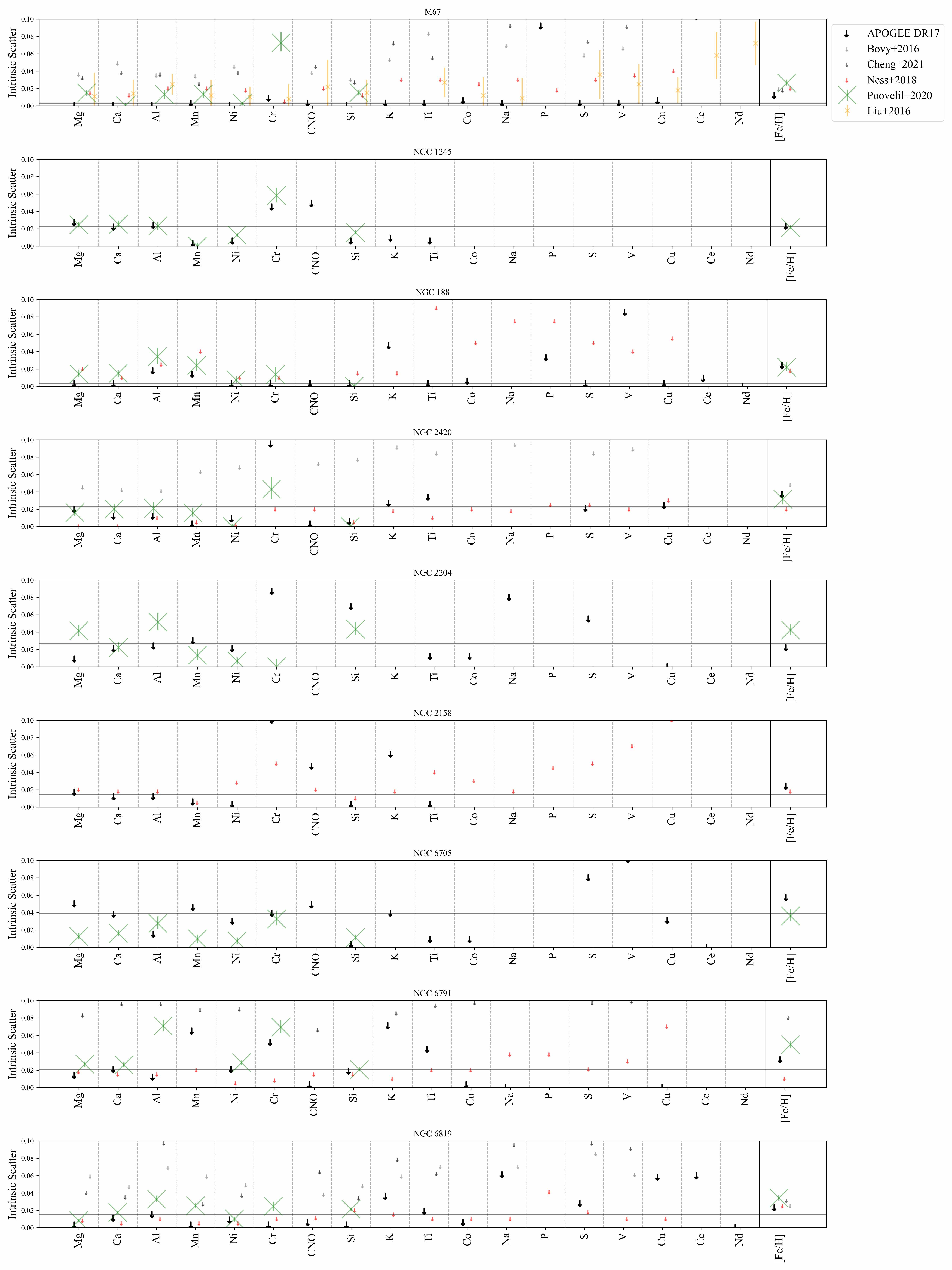}
  \caption{The literature comparison plots for the well studied OCs in our sample using APOGEE DR17 (colored in black). The horizontal line indicates the median 99.7$\%$ confidence limit for that cluster's homogeneity.}
\end{figure*}
\clearpage

\section{Open Cluster Sample Membership}
\label{sec:appendix_b}
Here we show the plots outlining the kinematic selection of our open cluster sample, as described in Section~\ref{sec:kinematics}.
\begin{figure}[hbt!]
    \centering
    \includegraphics[width=0.3\textwidth]{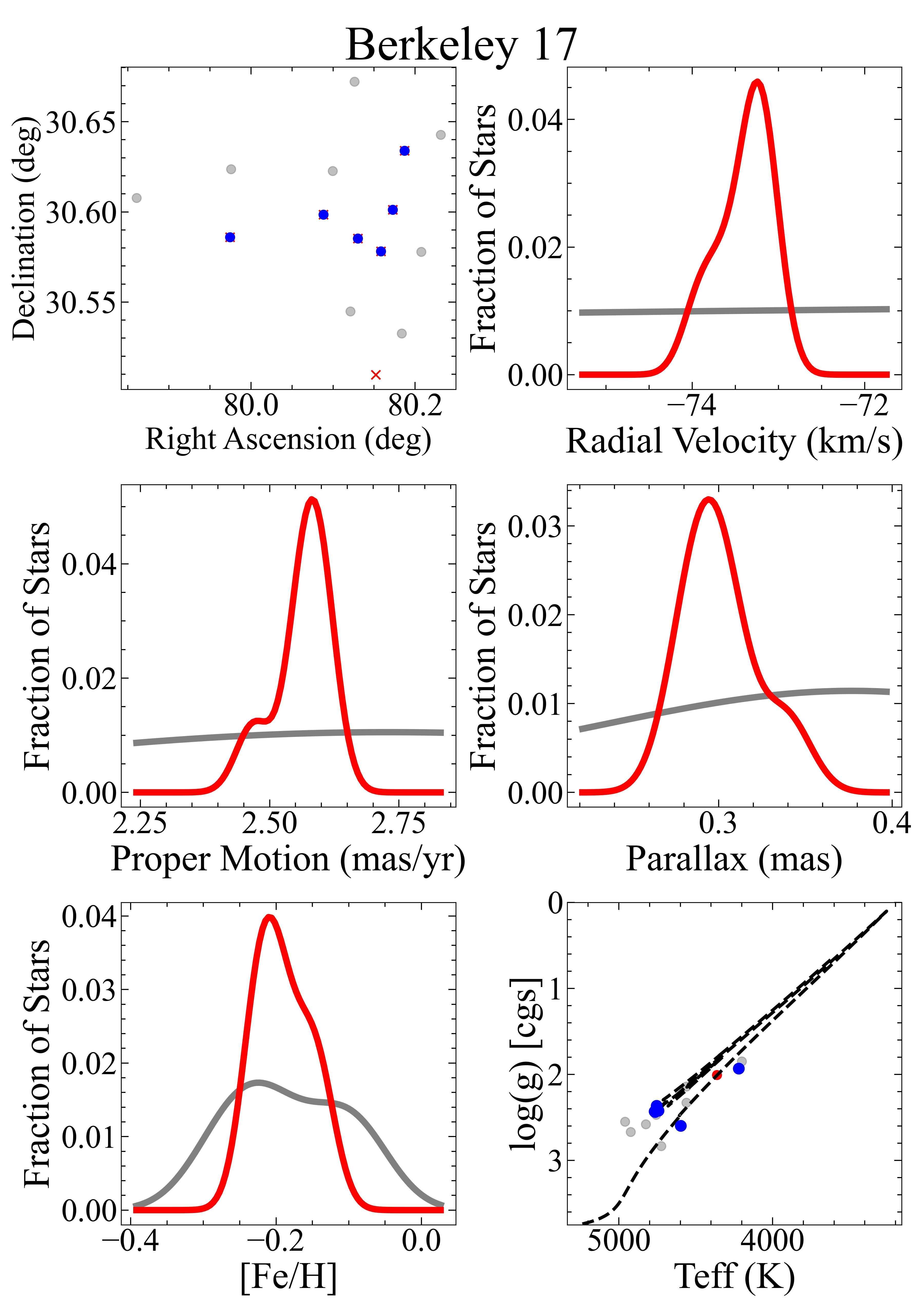}
    \includegraphics[width=0.3\textwidth]{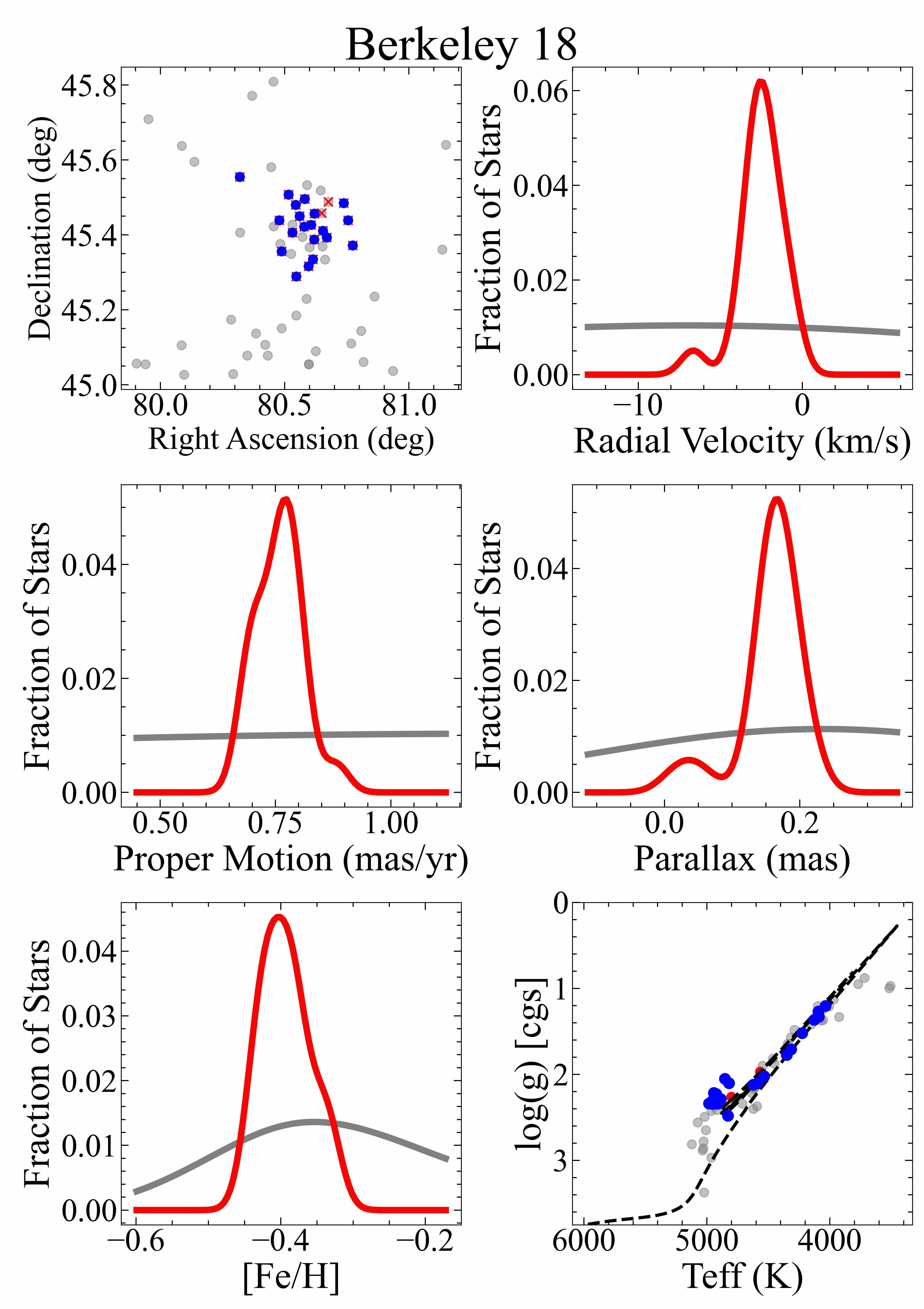}
    \includegraphics[width=0.3\textwidth]{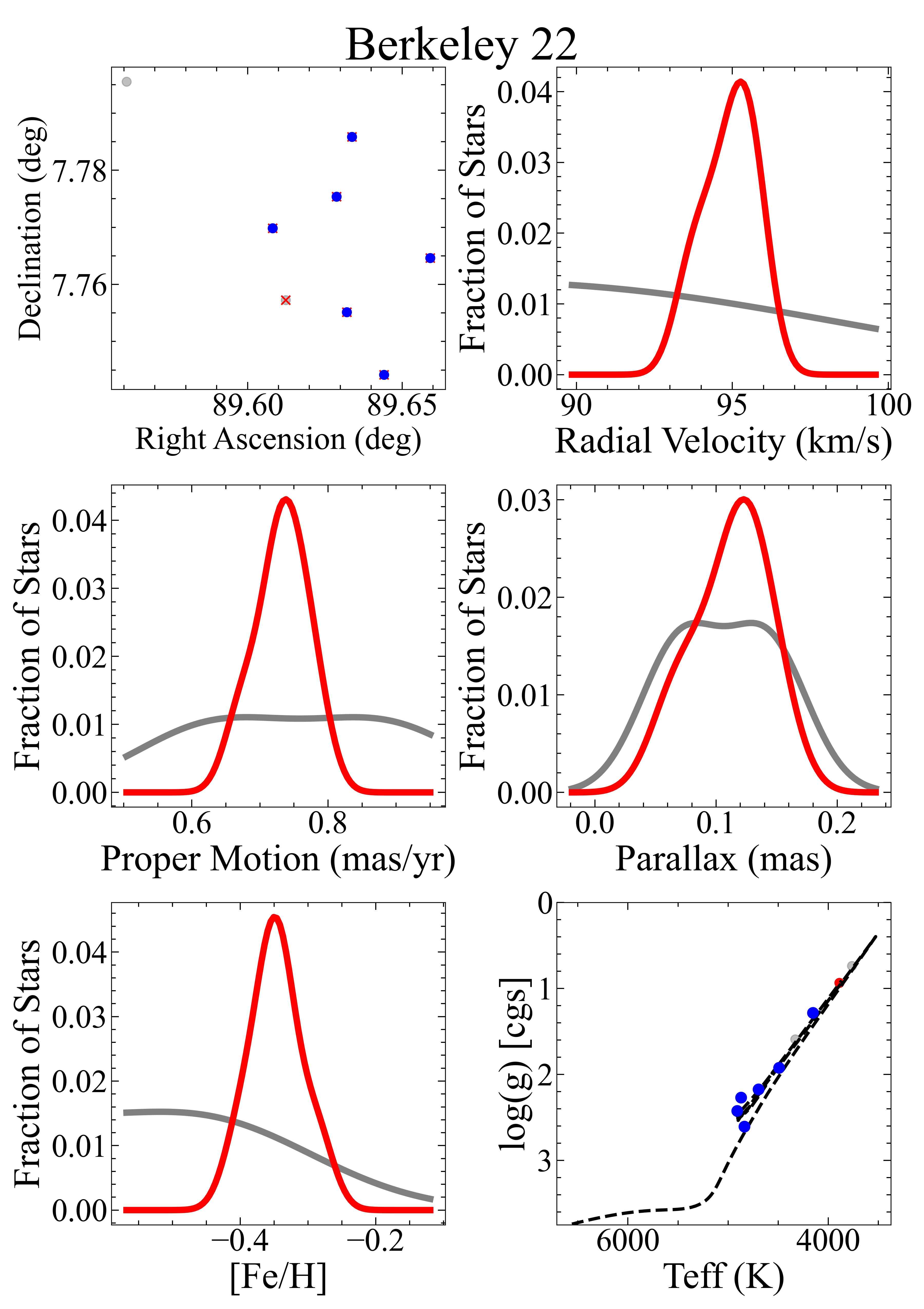}
    \includegraphics[width=0.3\textwidth]{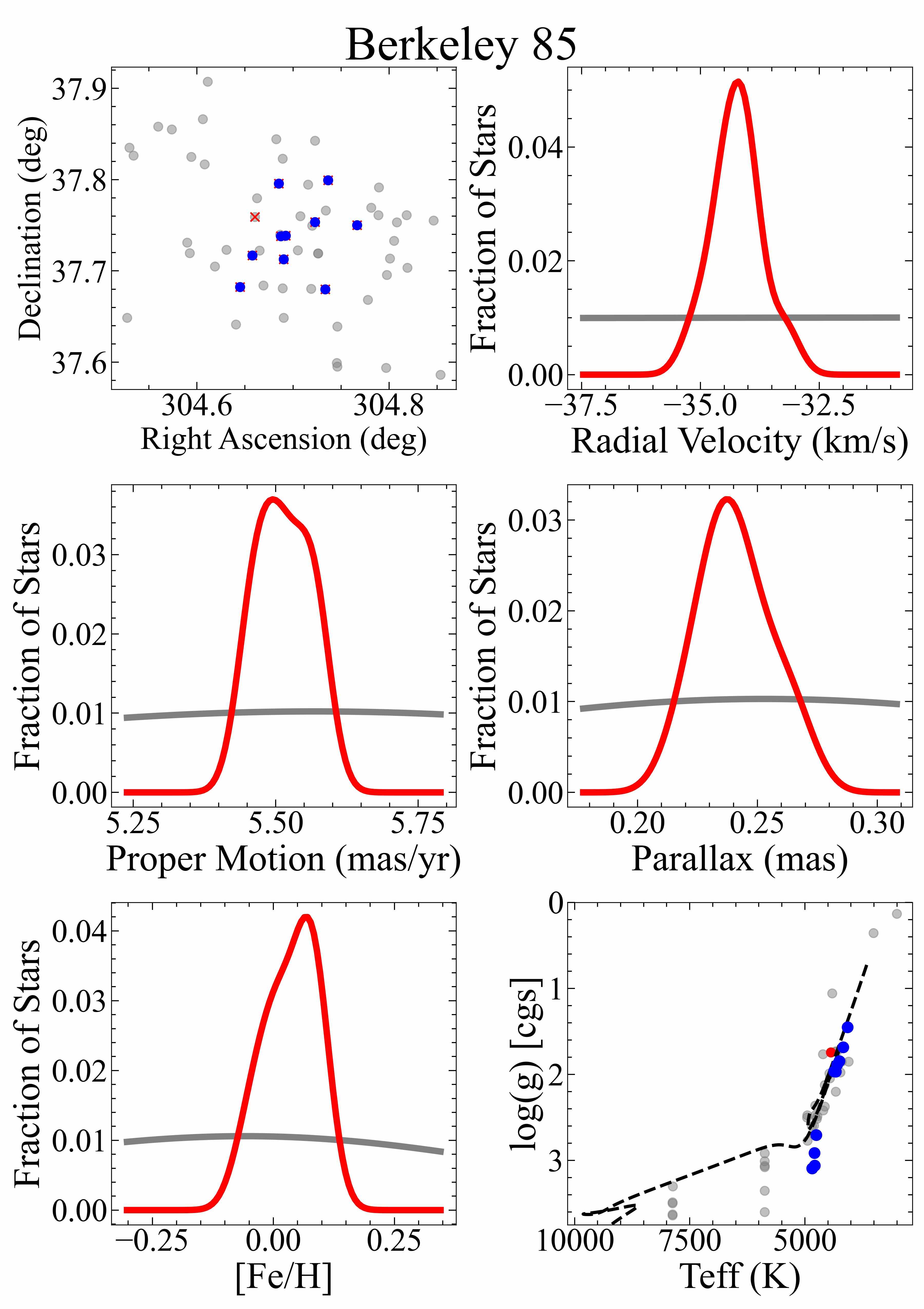}
    \includegraphics[width=0.3\textwidth]{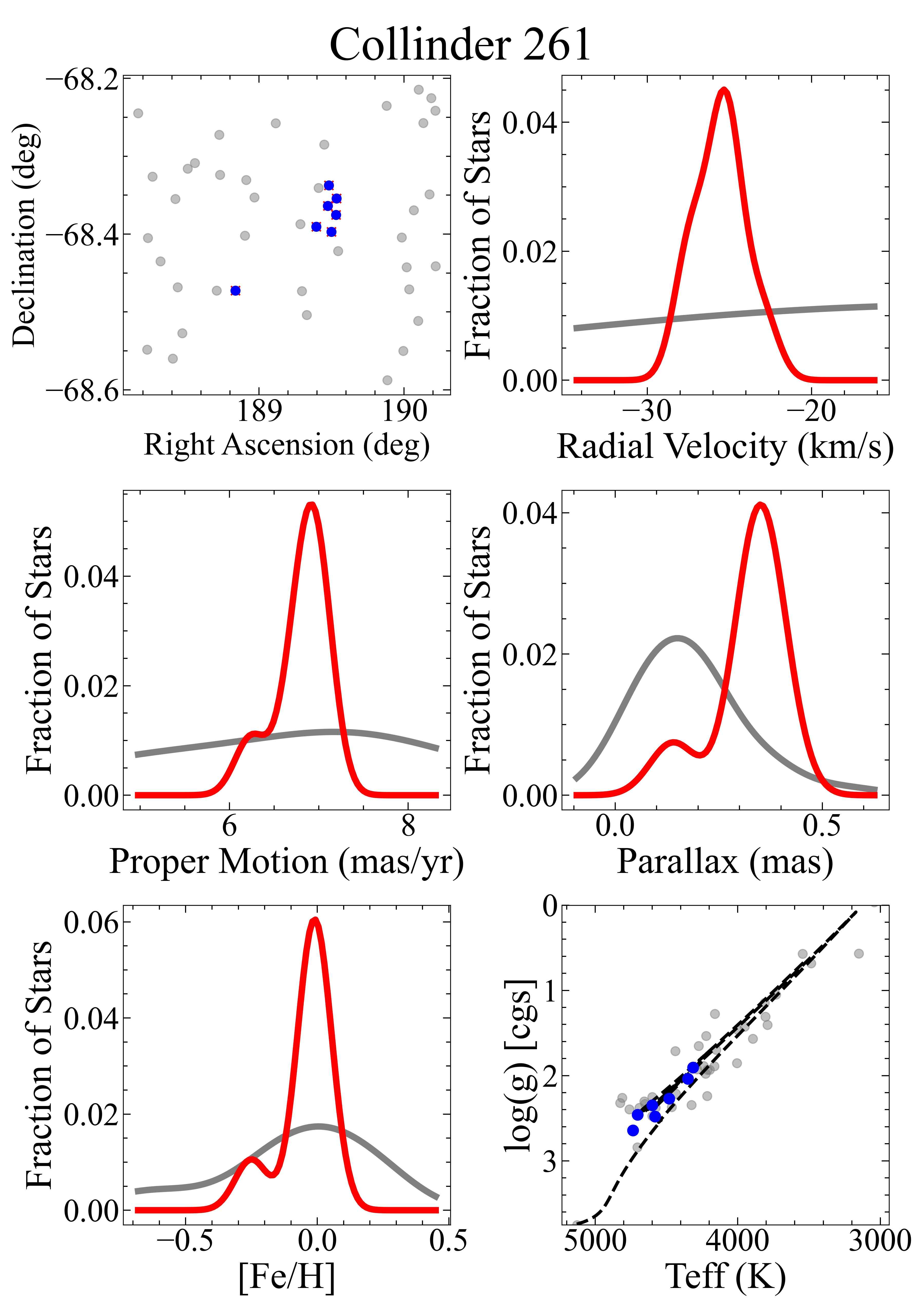}
    \includegraphics[width=0.3\textwidth]{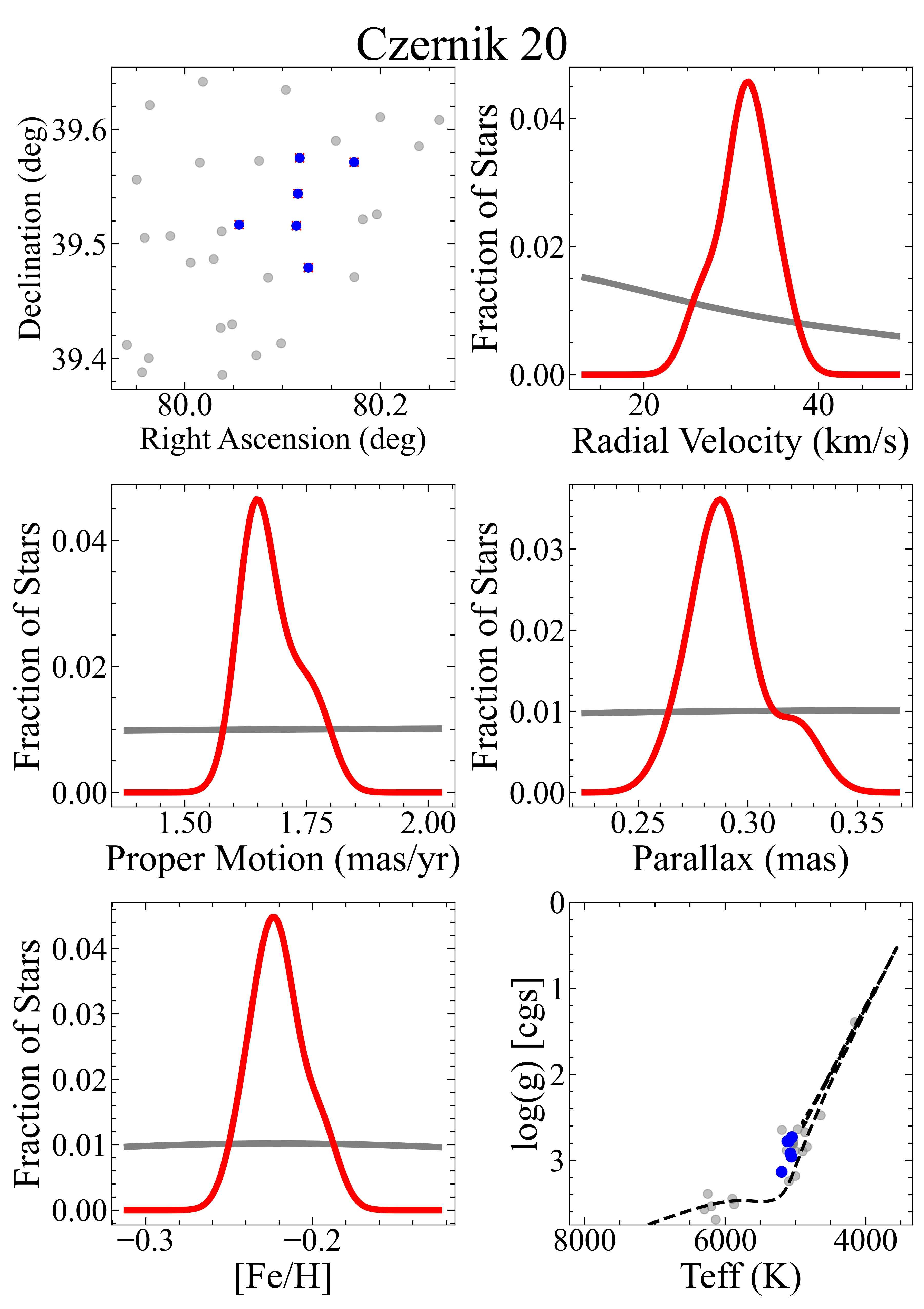}
    \includegraphics[width=0.3\textwidth]{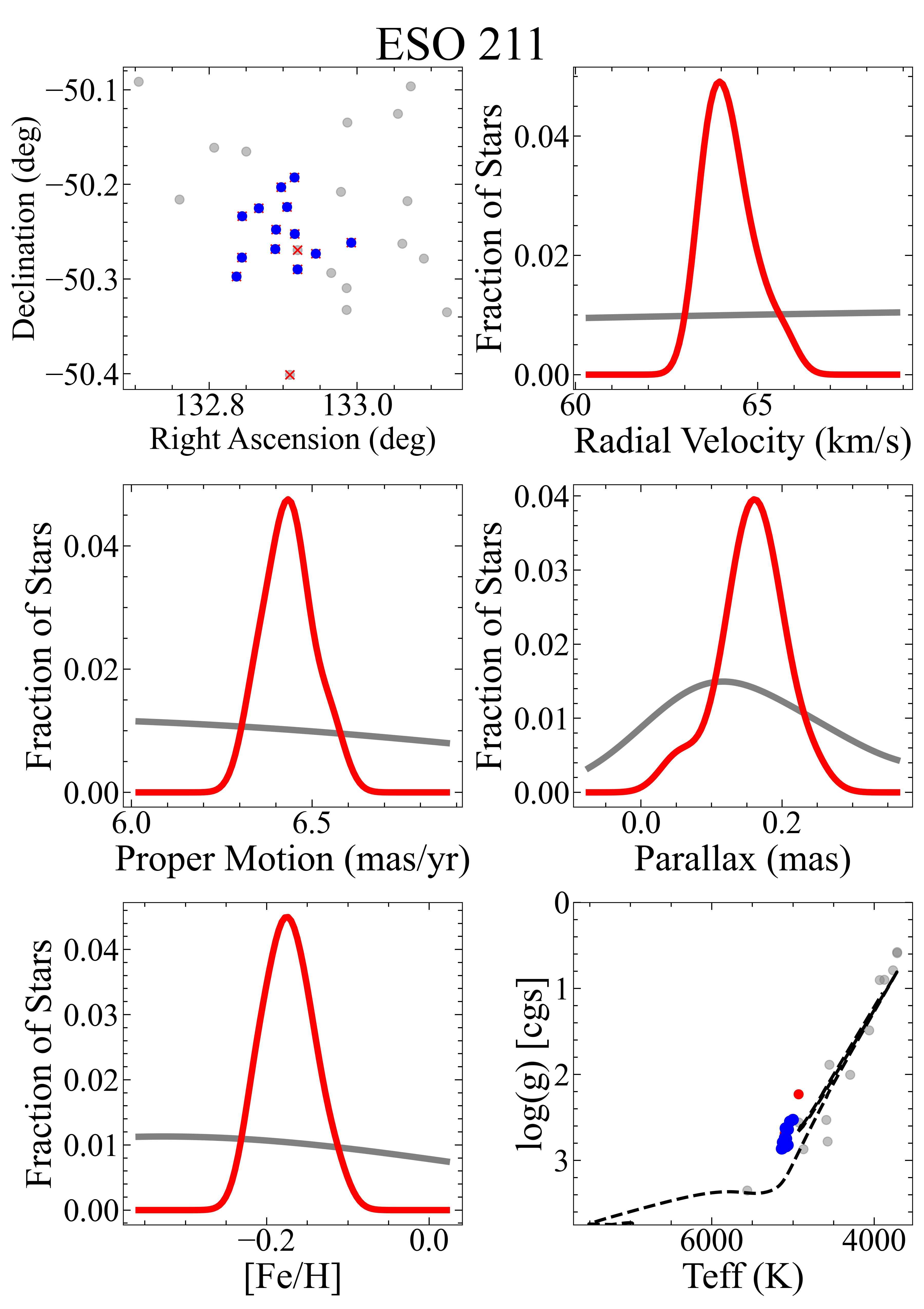}
    \includegraphics[width=0.3\textwidth]{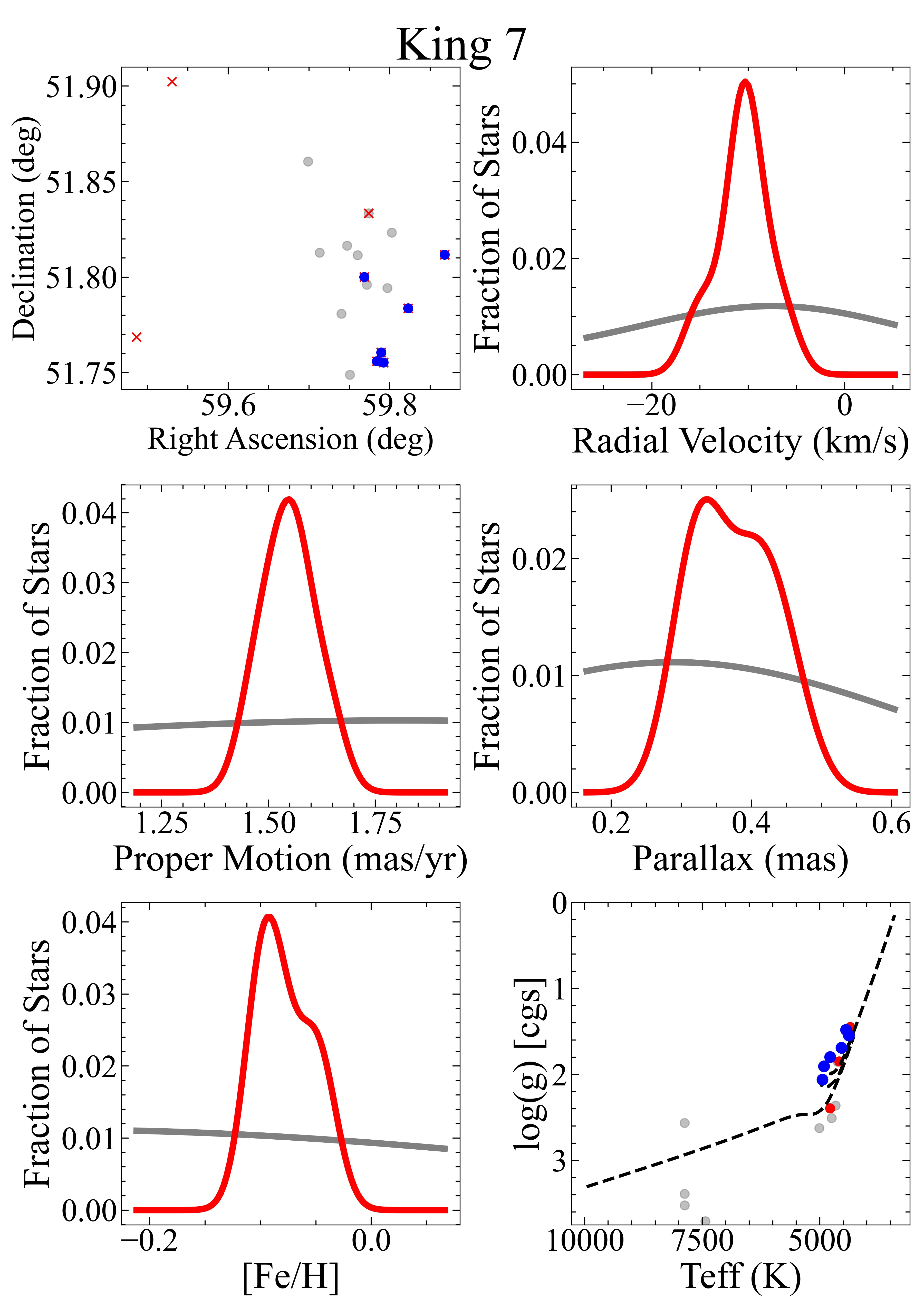}
    \includegraphics[width=0.3\textwidth]{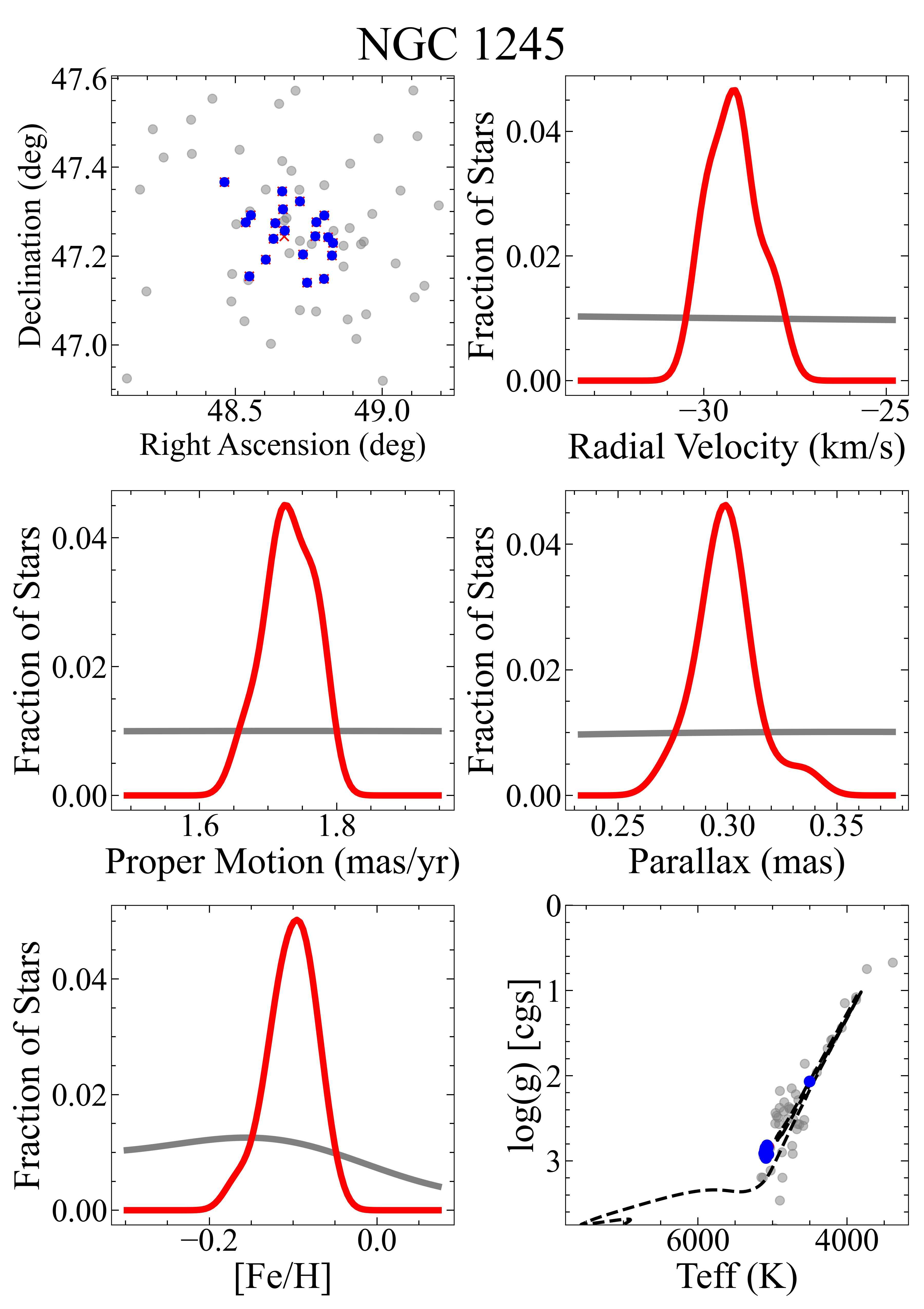}
    \caption{These figures follow the same layout as Figure \ref{fig:kinematic} in Section \ref{sec:kinematics}.}
\end{figure}
\begin{figure}[hbt!]
    \centering
    \includegraphics[width=0.3\textwidth]{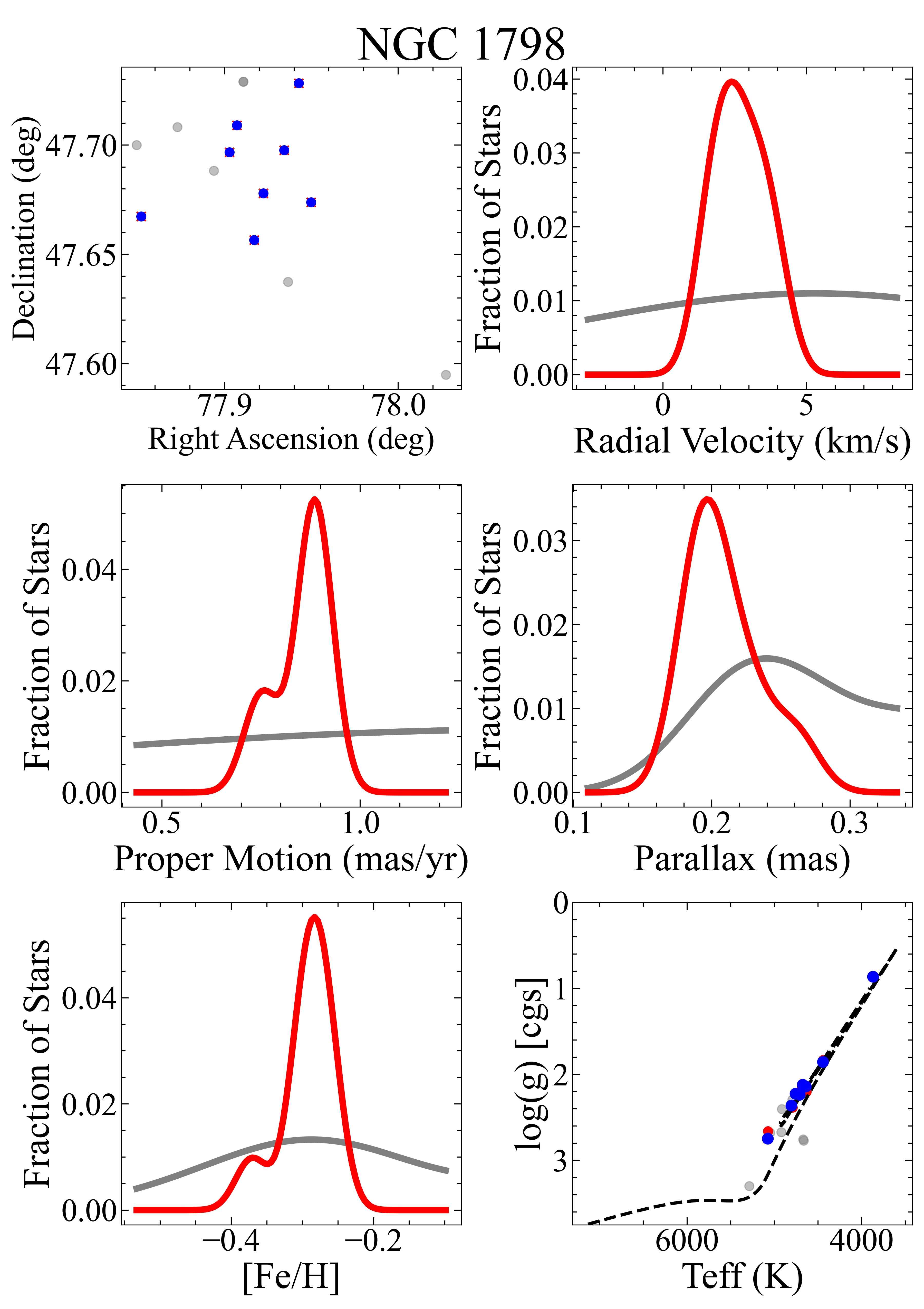}
    \includegraphics[width=0.3\textwidth]{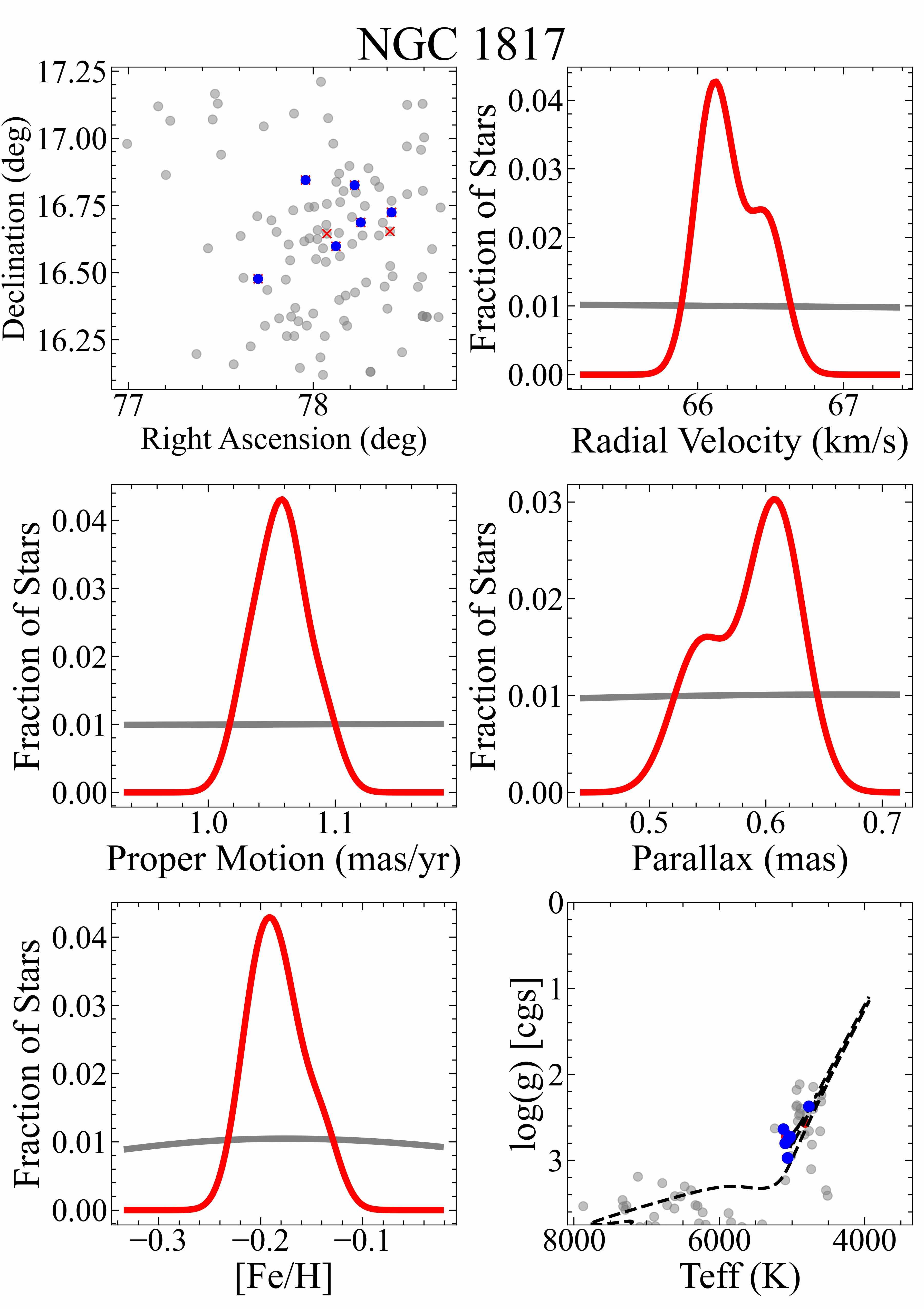}
    \includegraphics[width=0.3\textwidth]{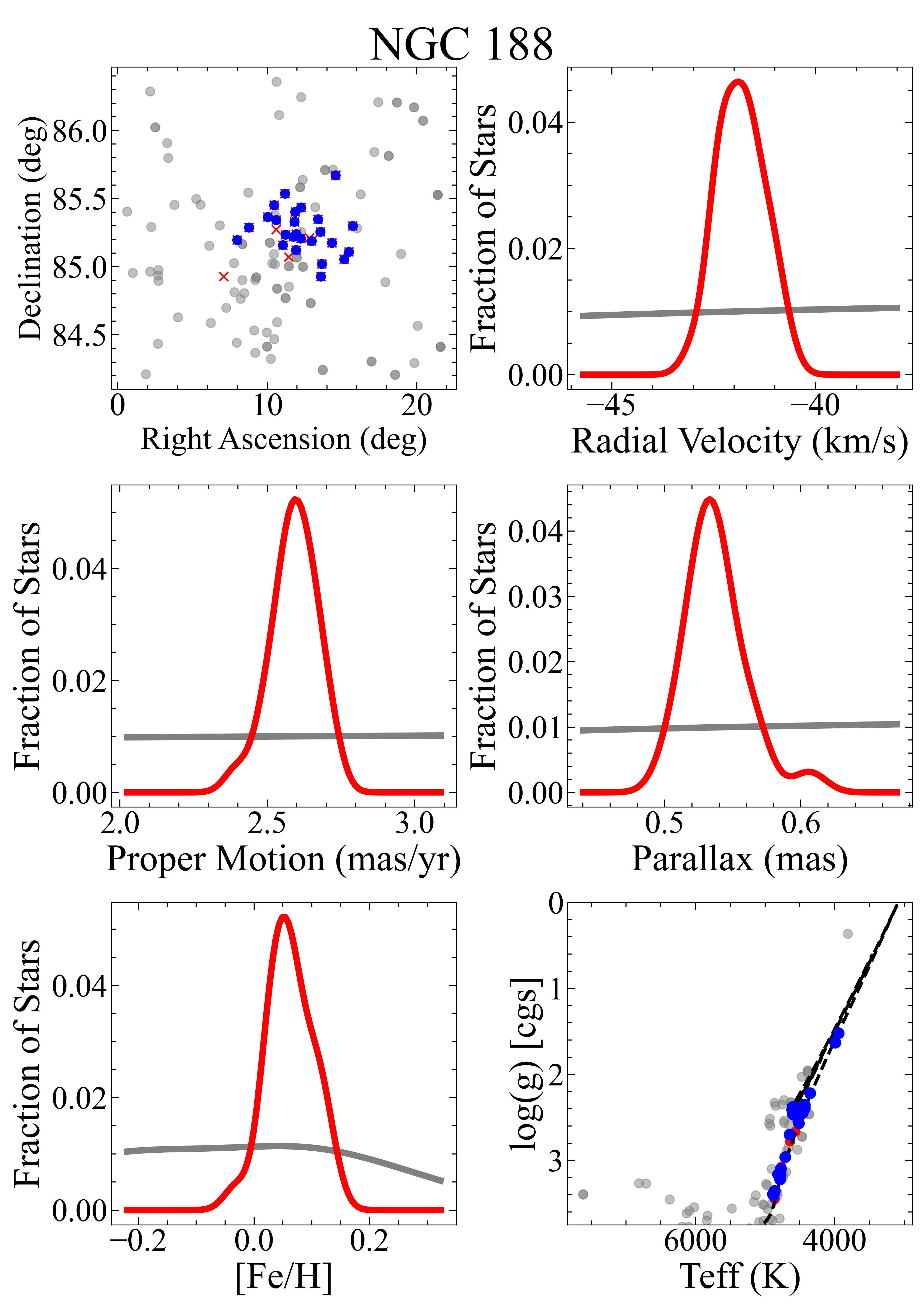}
    \includegraphics[width=0.3\textwidth]{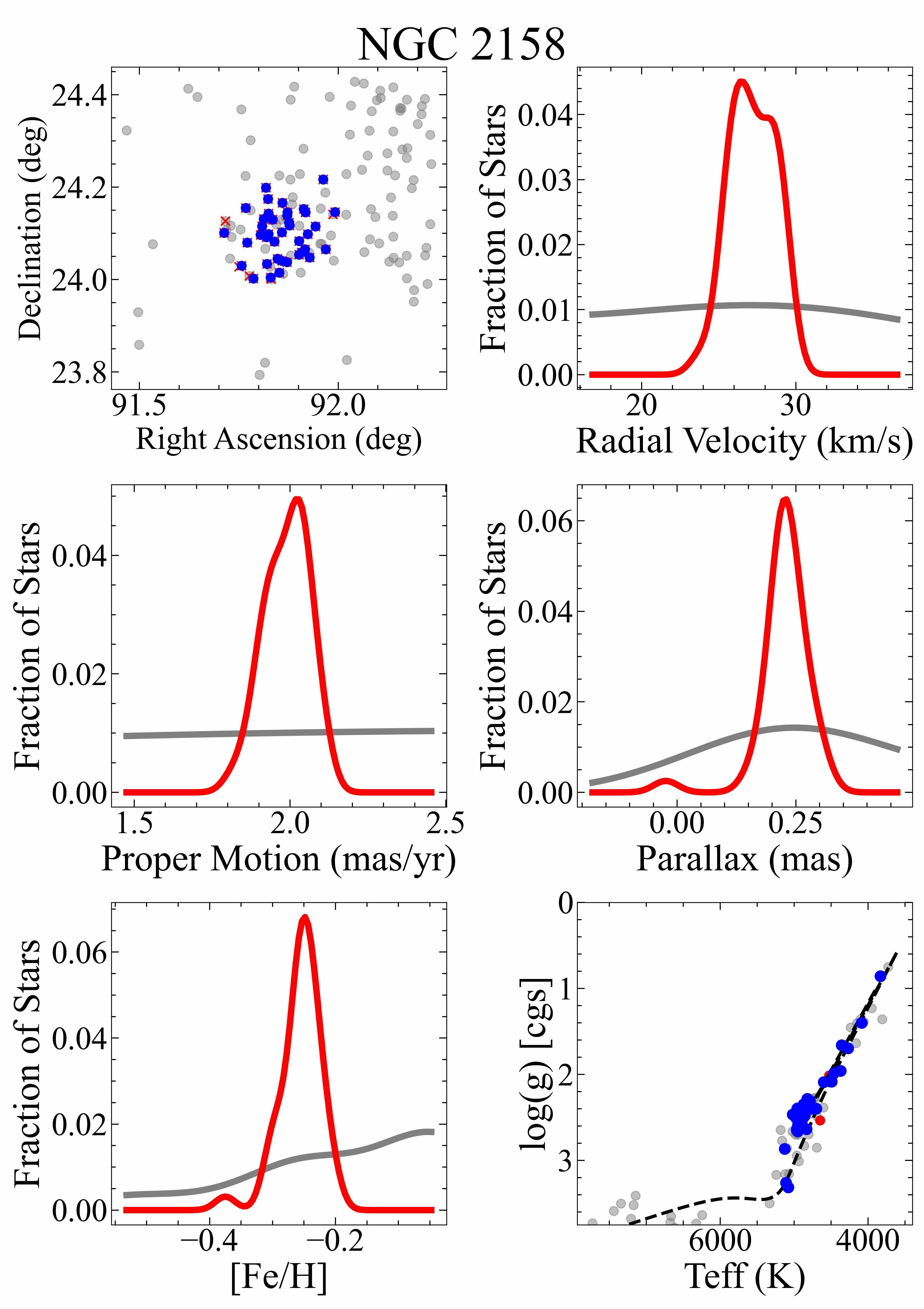}
    \includegraphics[width=0.3\textwidth]{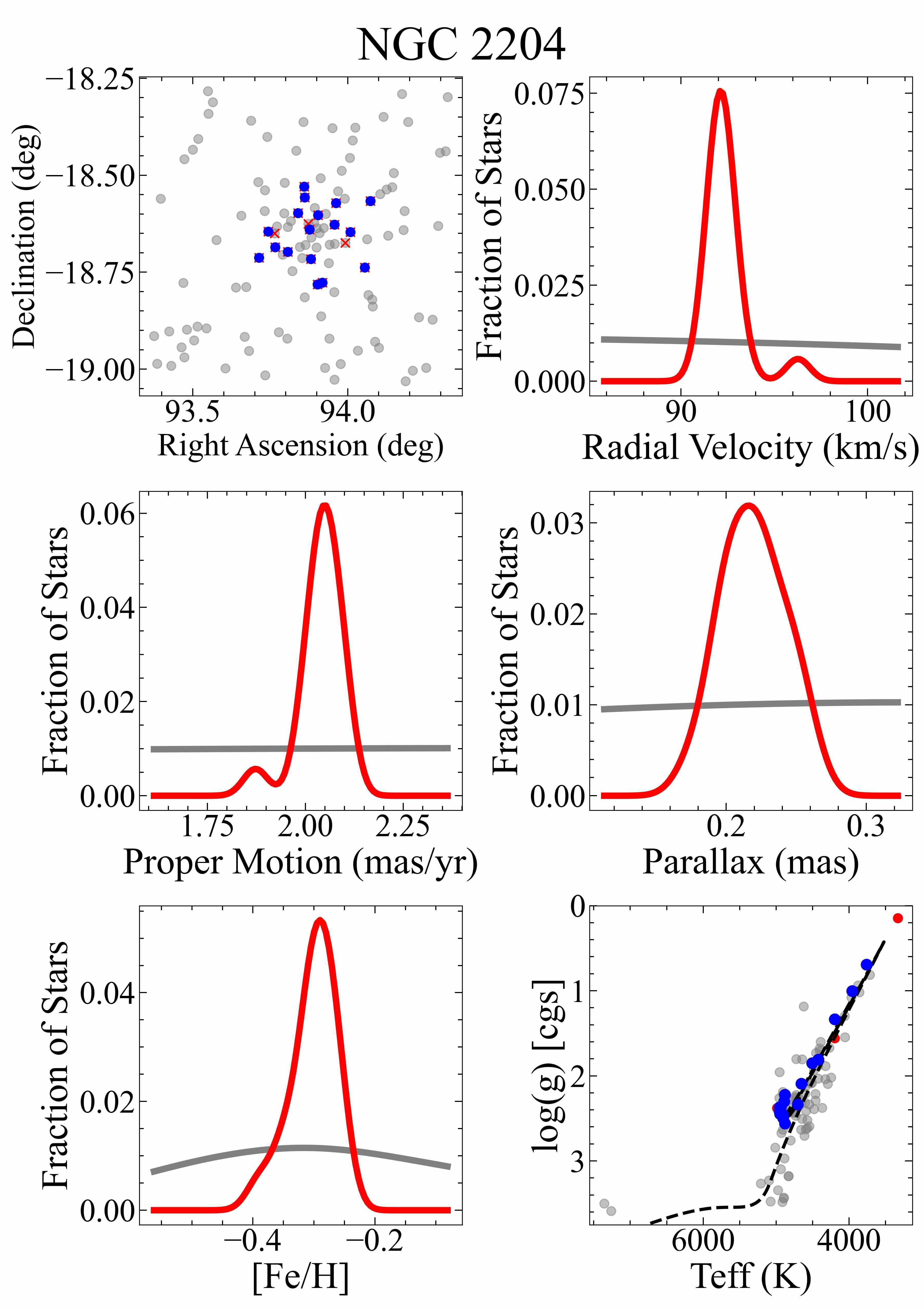}
    \includegraphics[width=0.3\textwidth]{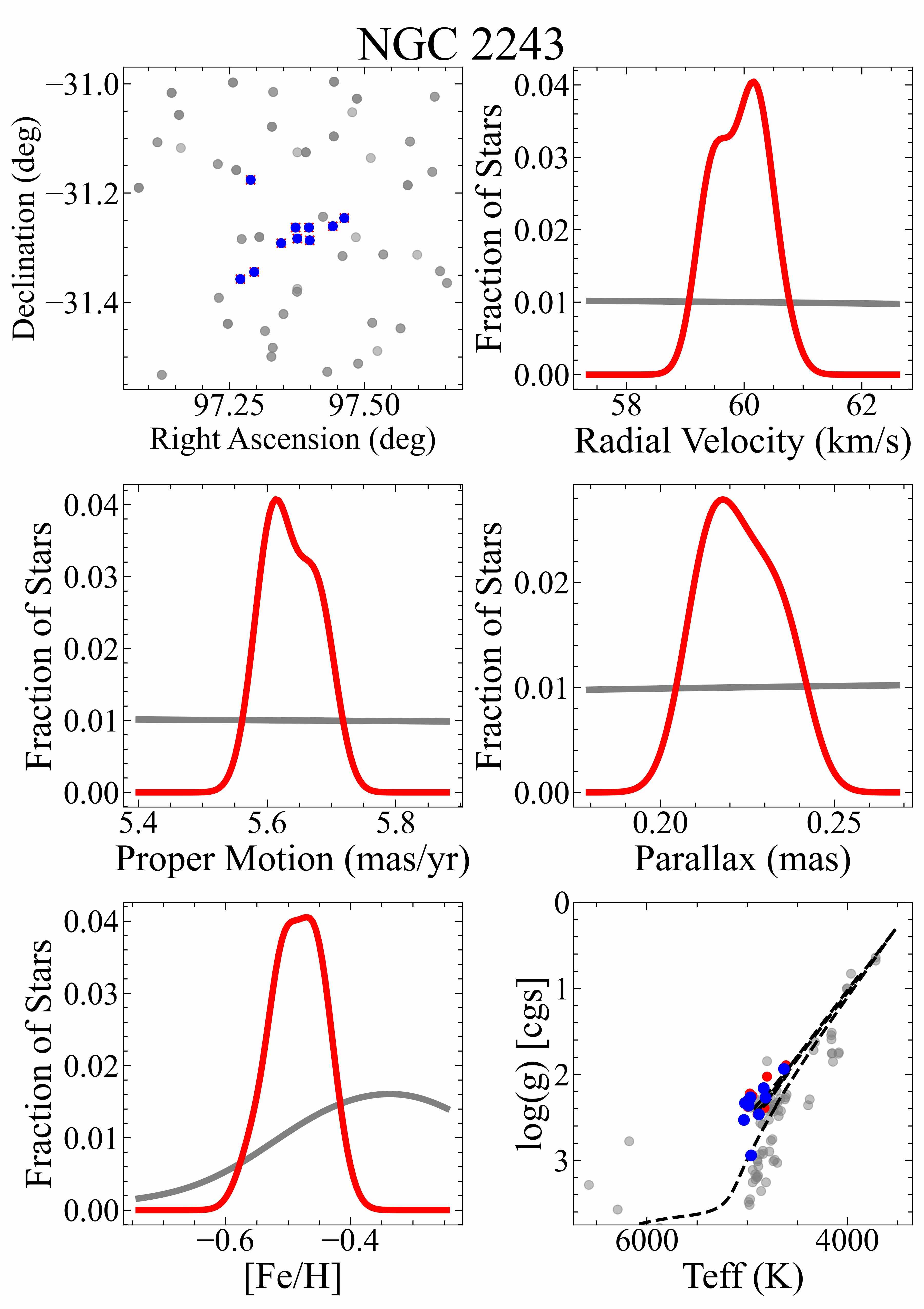}
    \includegraphics[width=0.3\textwidth]{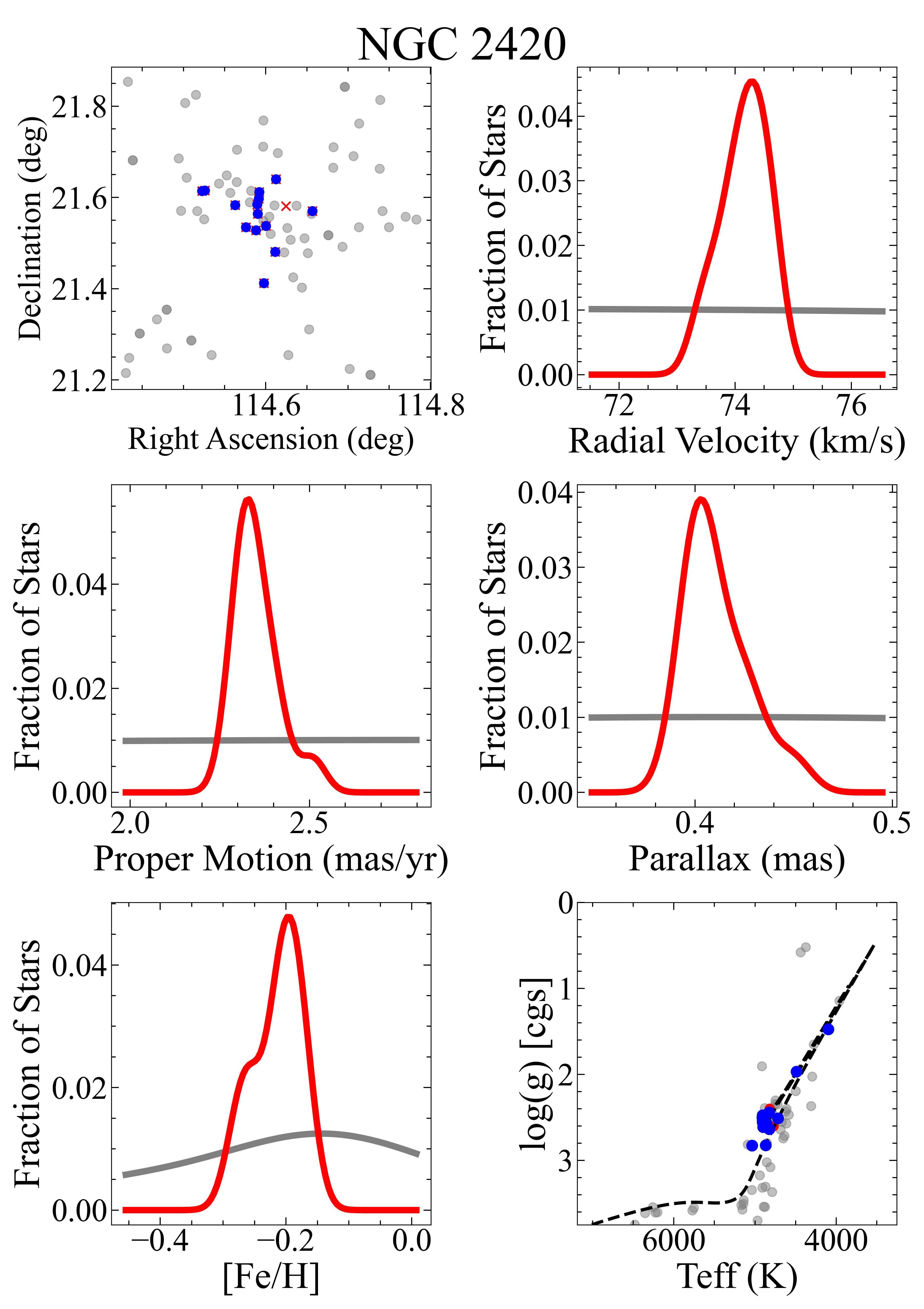}
    \includegraphics[width=0.3\textwidth]{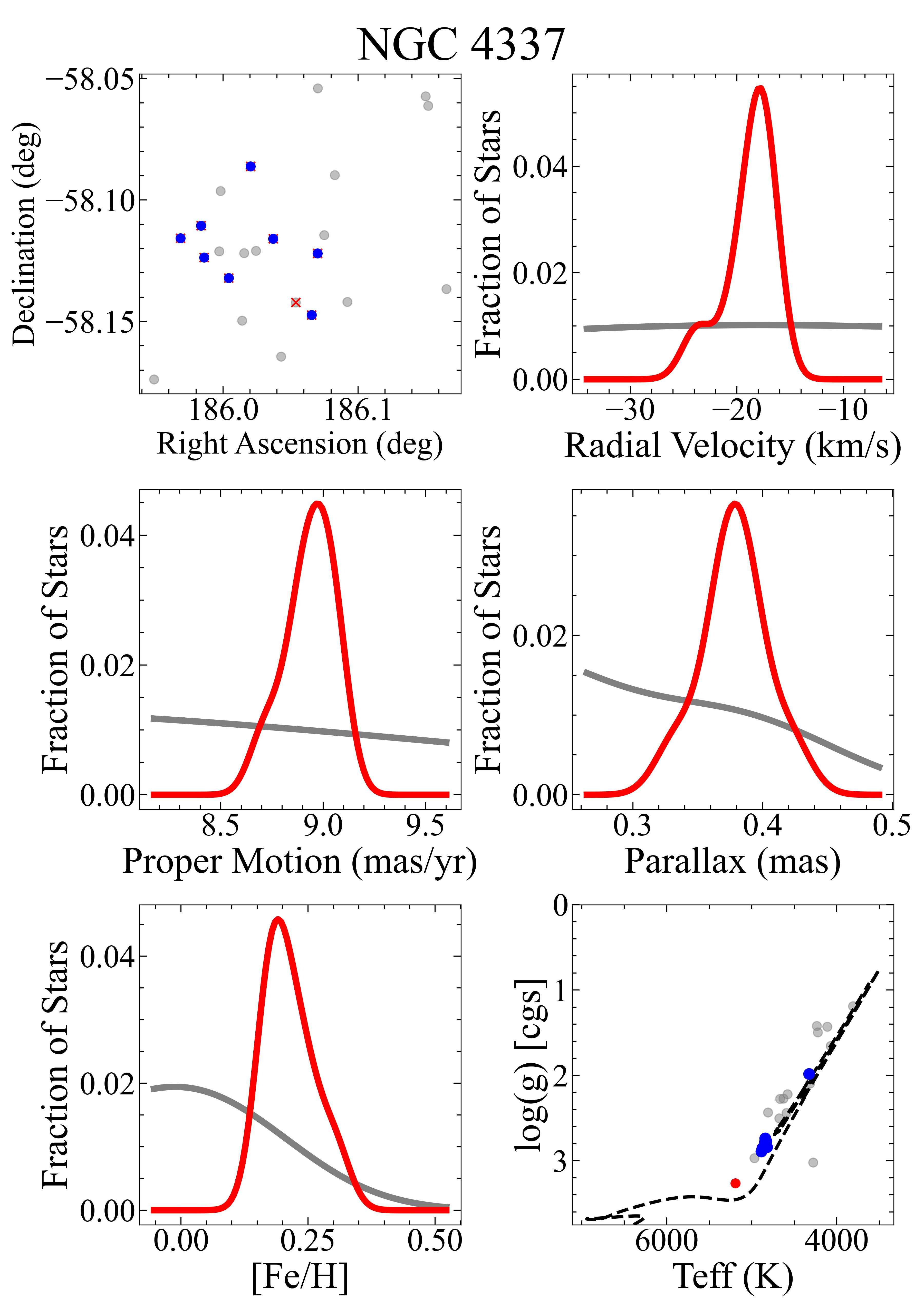}
    \includegraphics[width=0.3\textwidth]{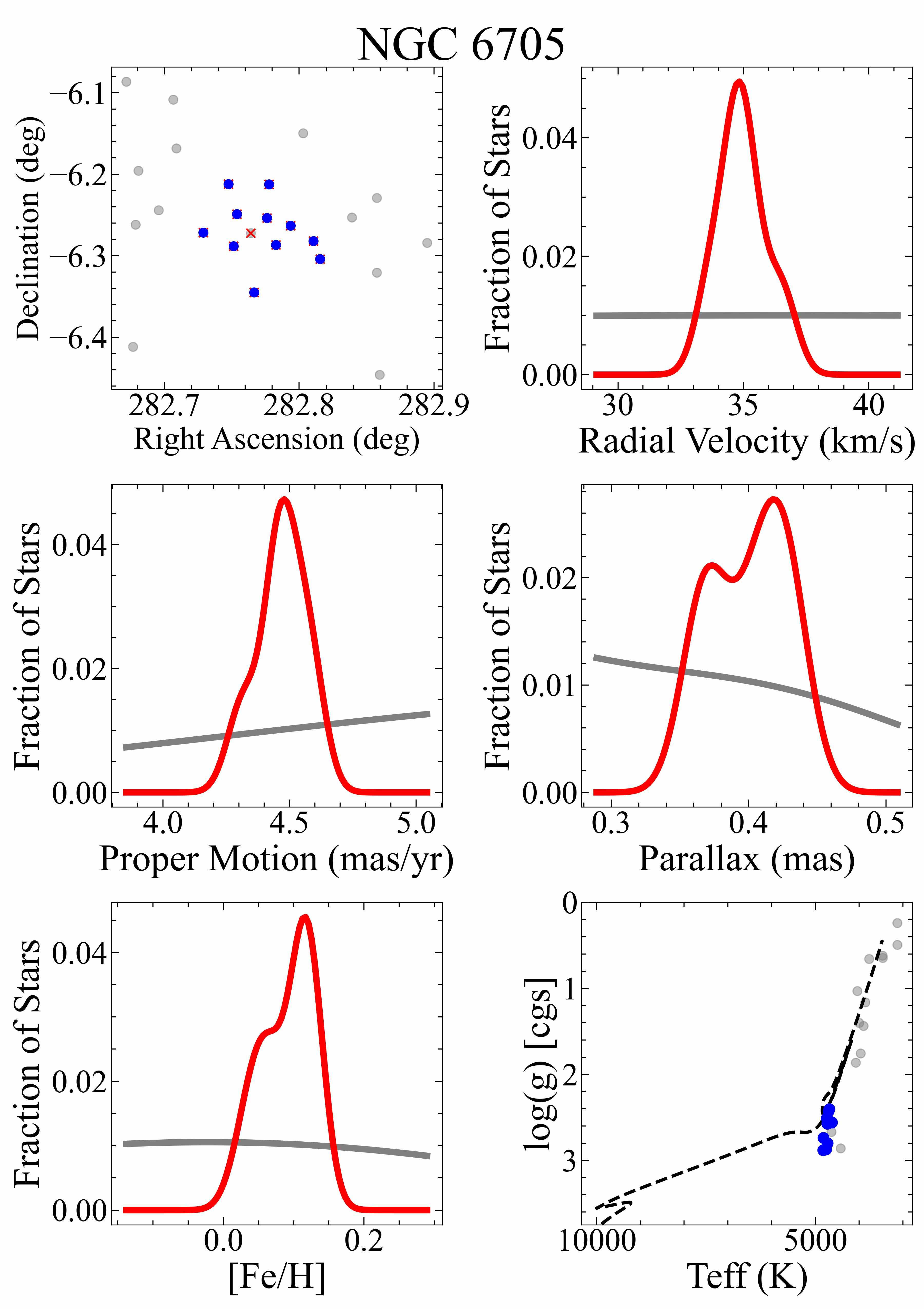}
    \caption{These figures follow the same layout as Figure \ref{fig:kinematic} in Section \ref{sec:kinematics}.}
\end{figure}
\begin{figure}
    \centering
    \includegraphics[width=0.3\textwidth]{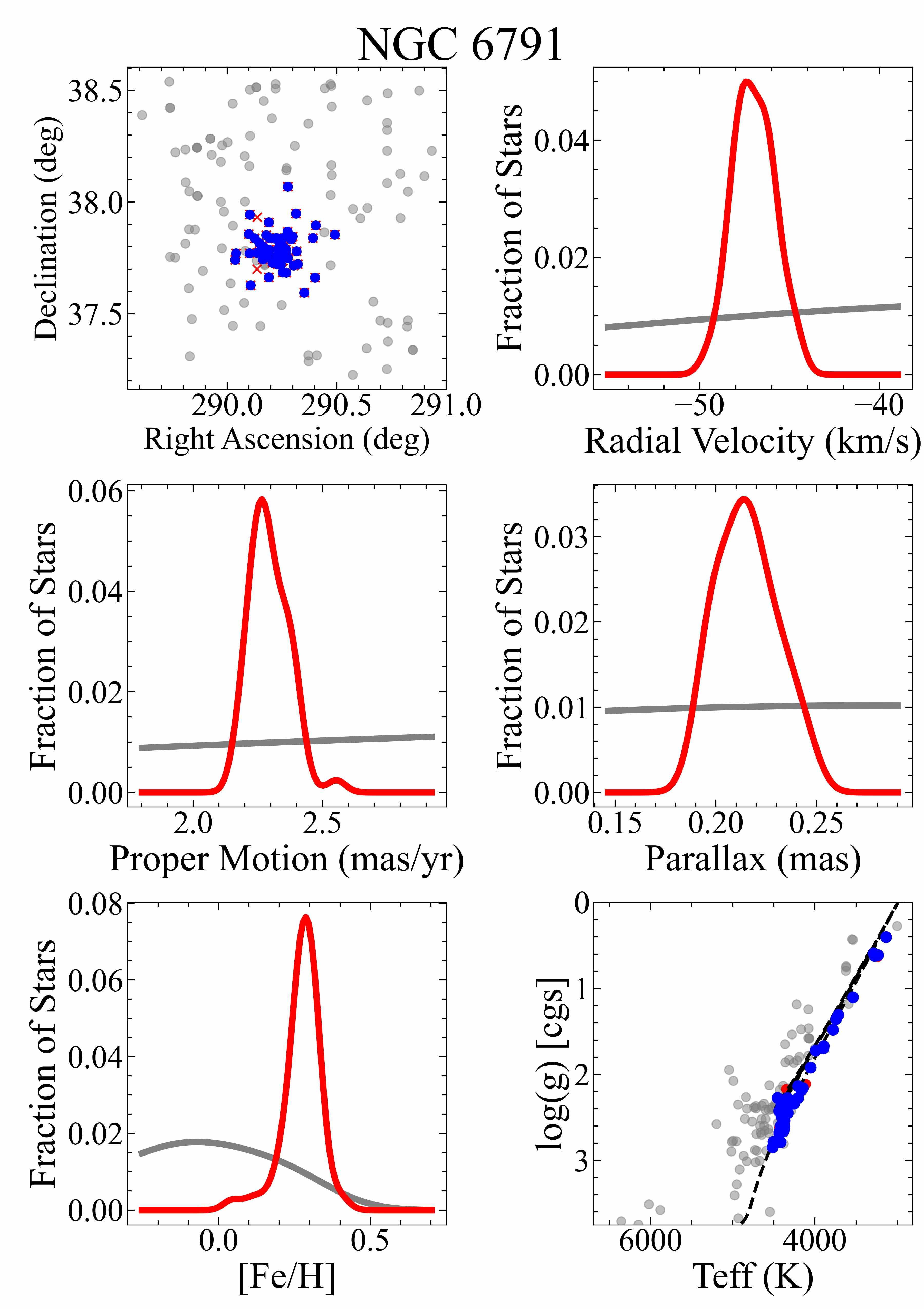}
    \includegraphics[width=0.3\textwidth]{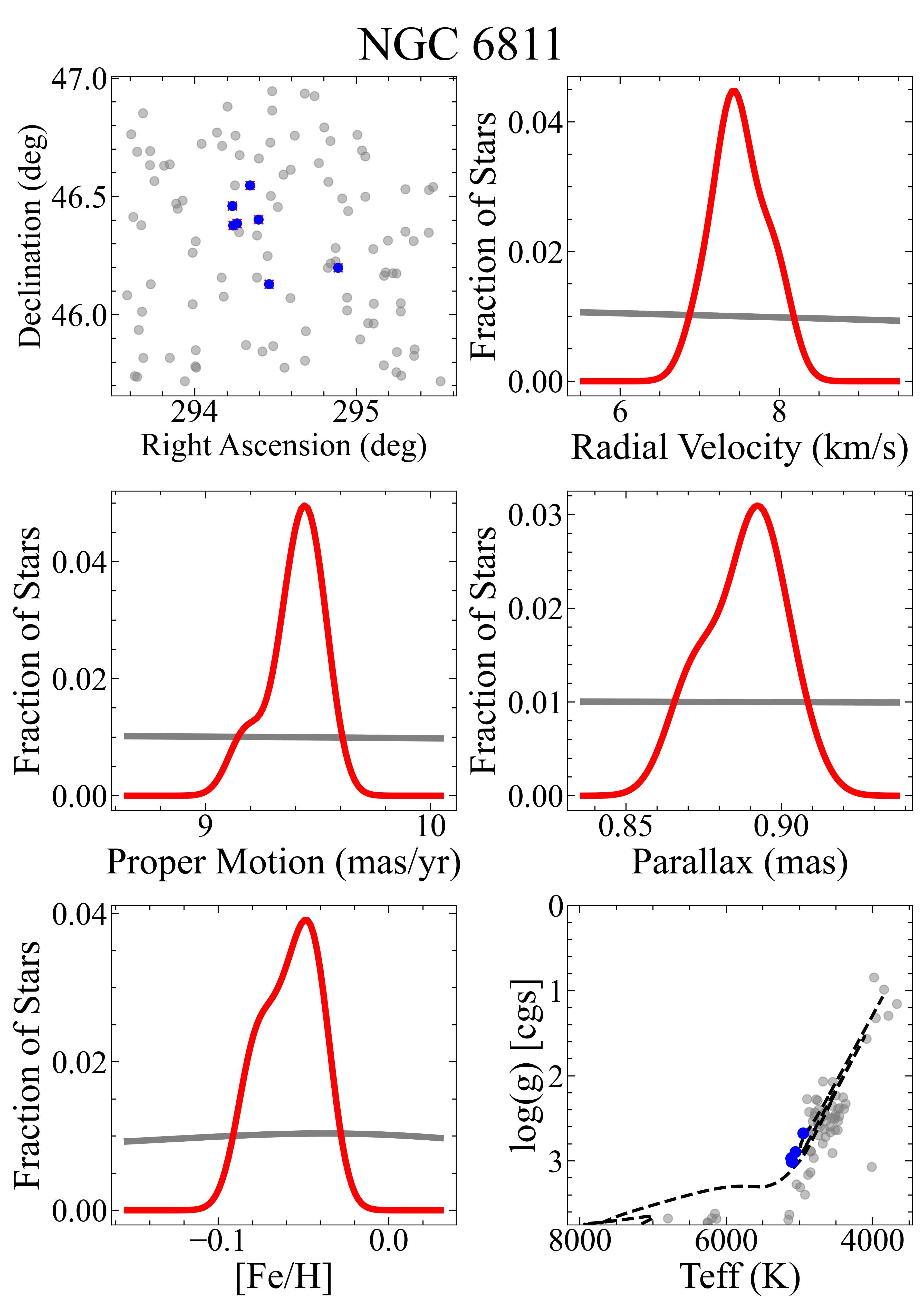}
    \includegraphics[width=0.3\textwidth]{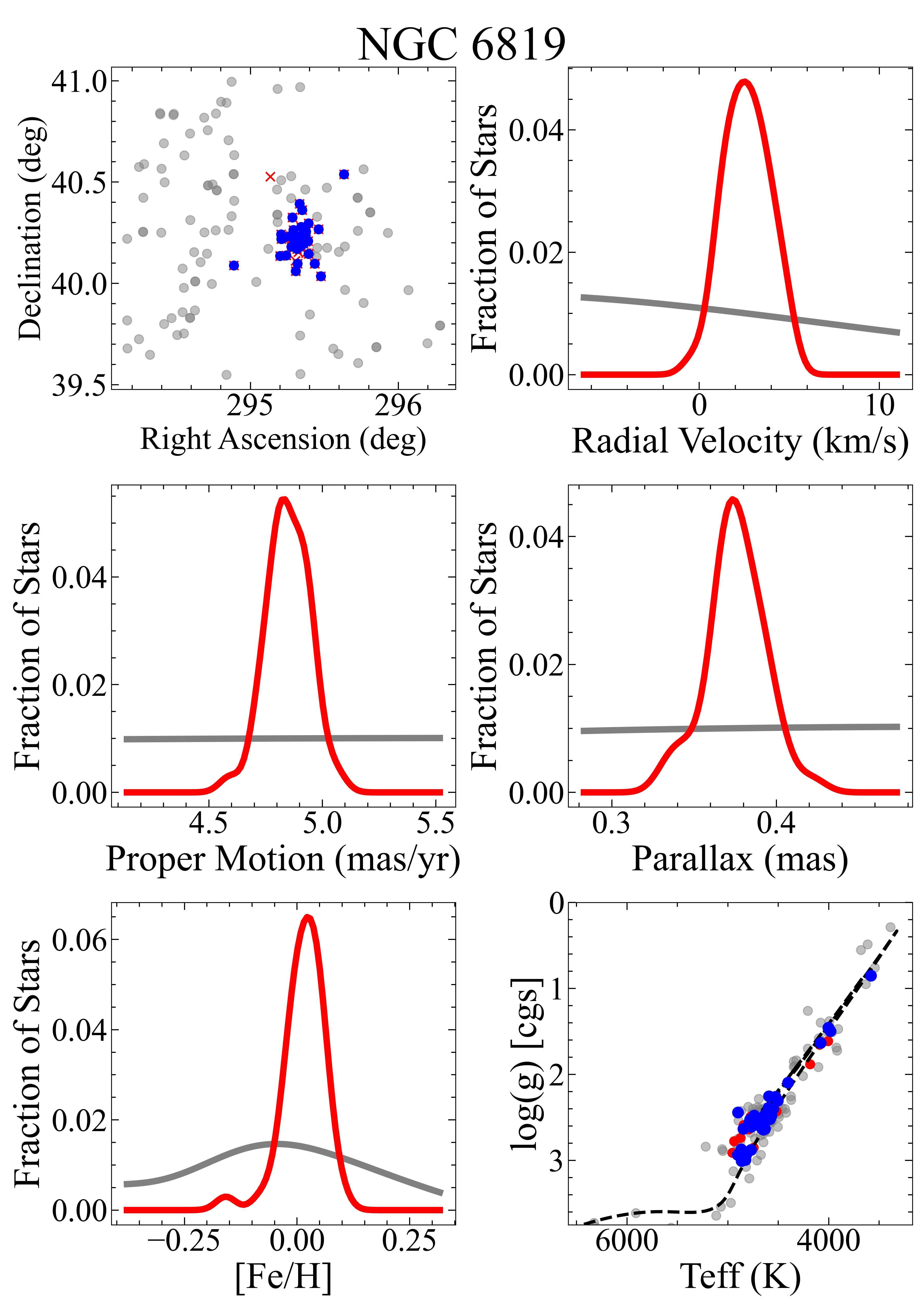}
    \includegraphics[width=0.3\textwidth]{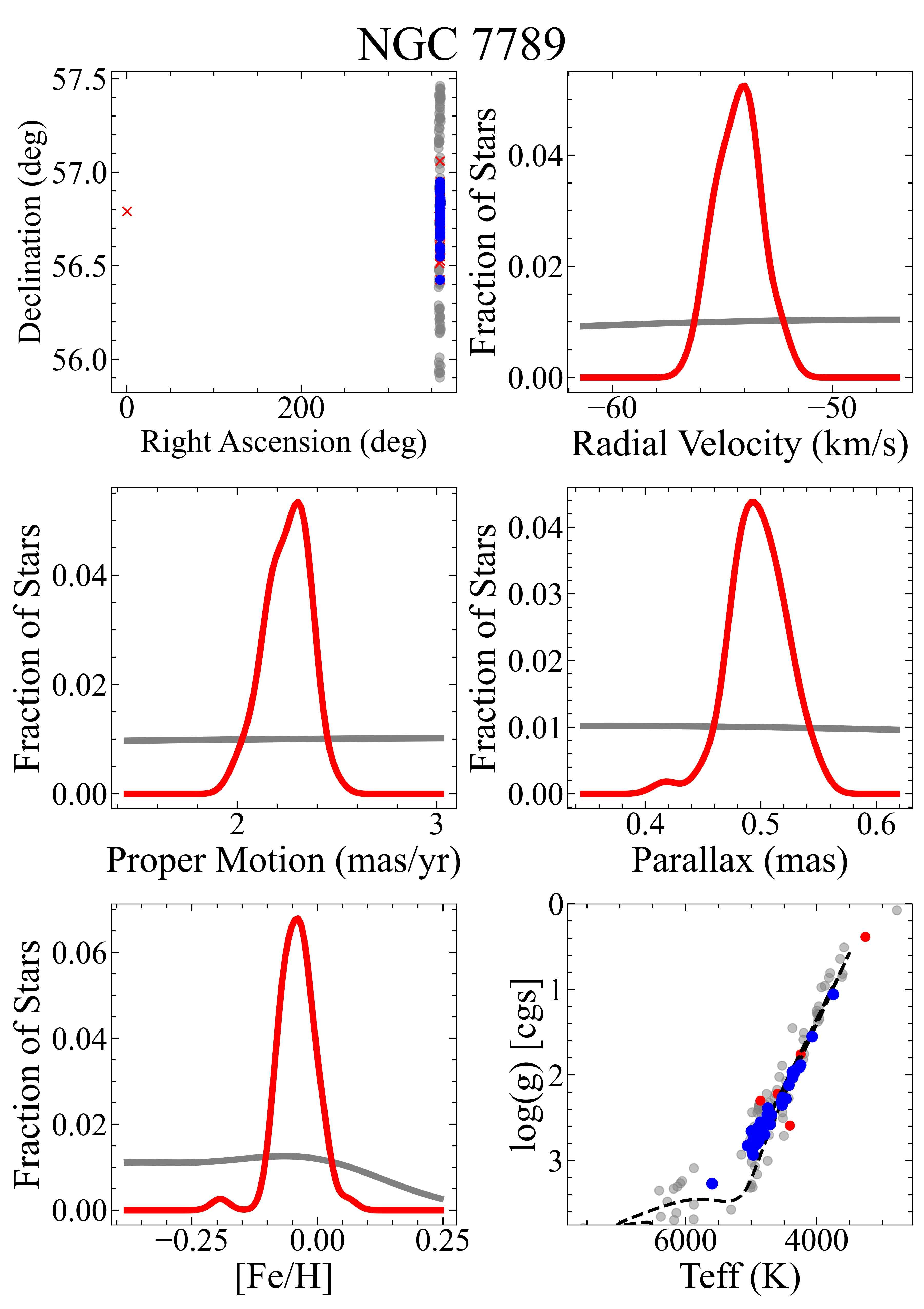}
    \includegraphics[width=0.3\textwidth]{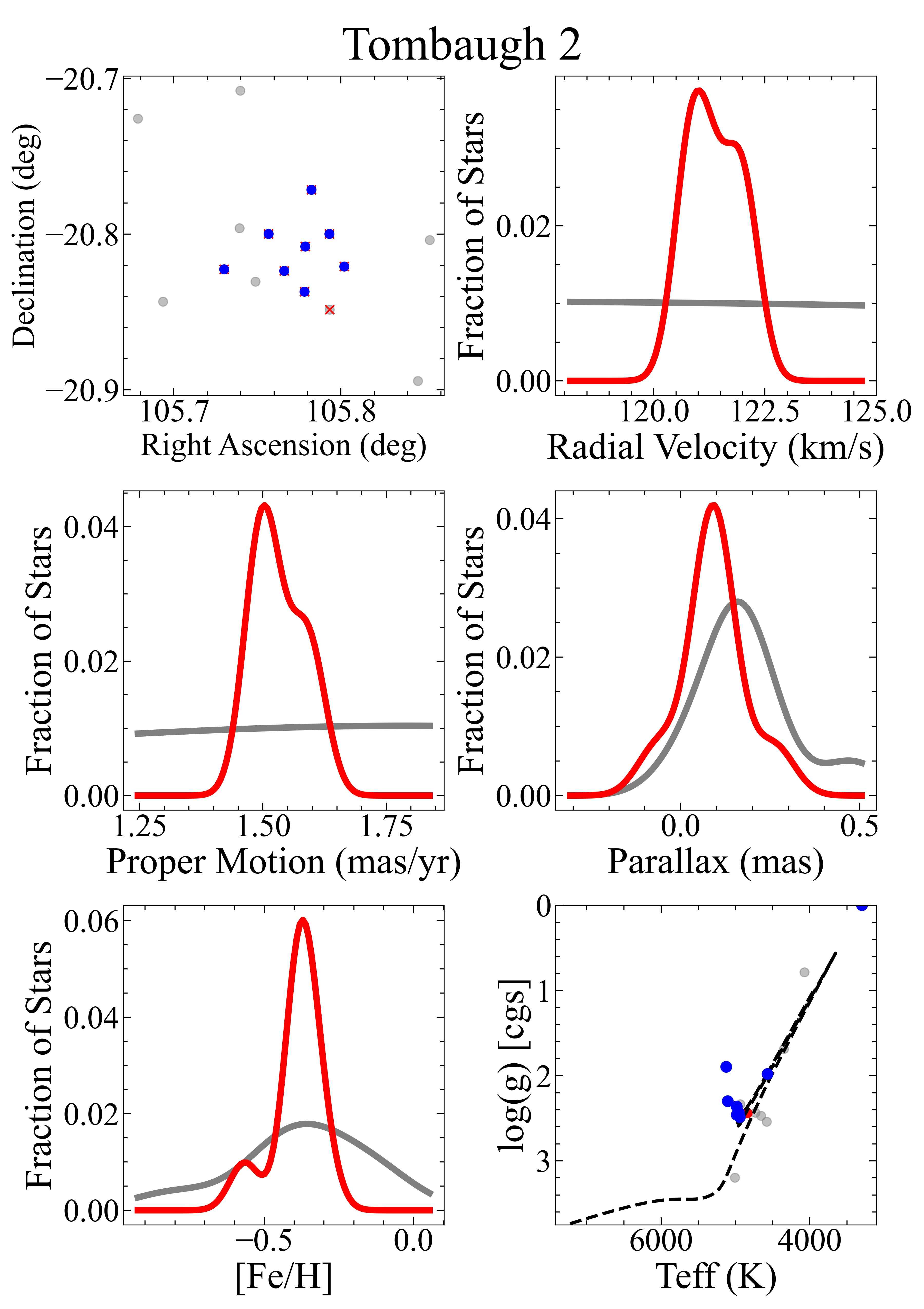}
    \includegraphics[width=0.3\textwidth]{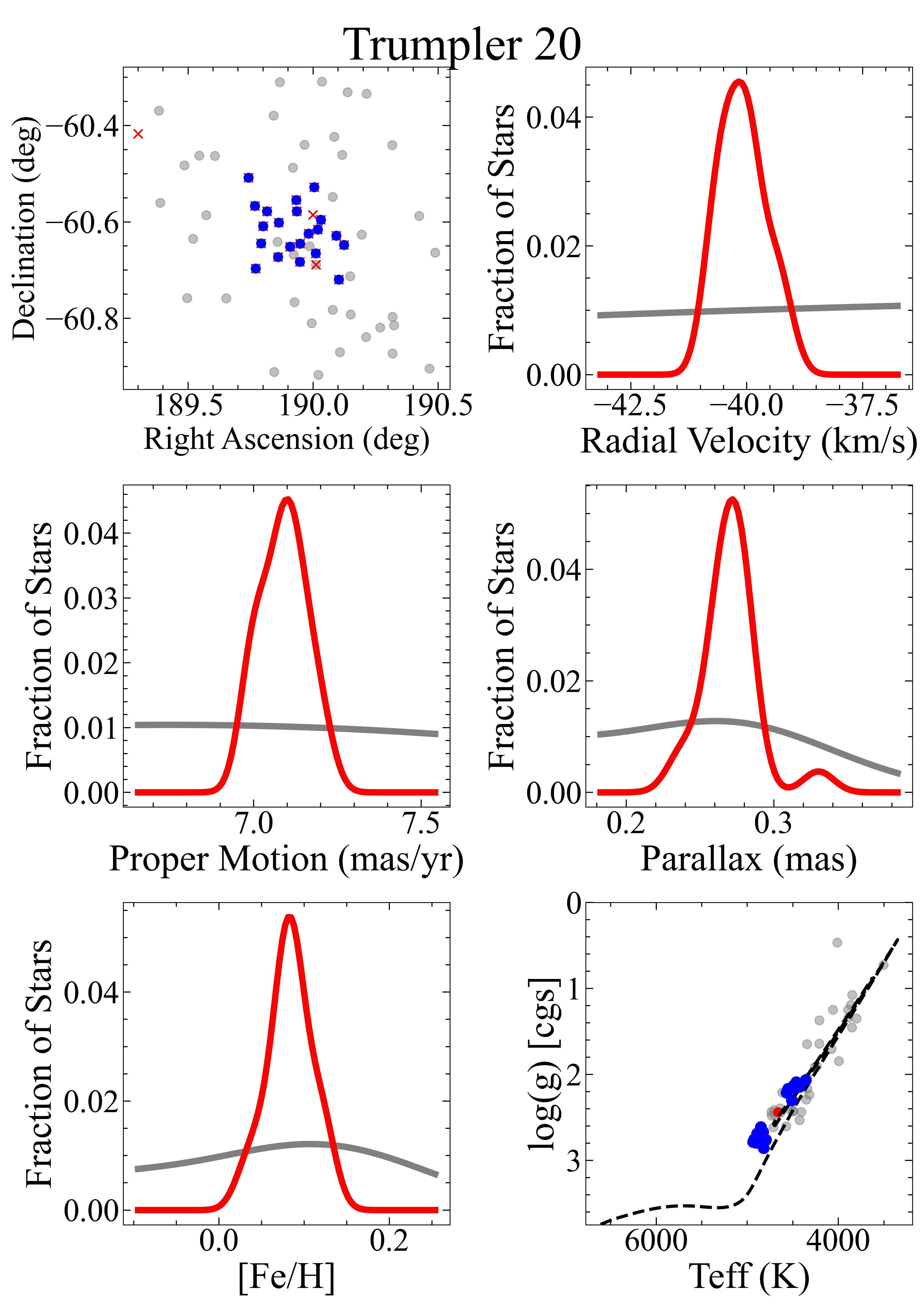}
    \includegraphics[width=0.3\textwidth]{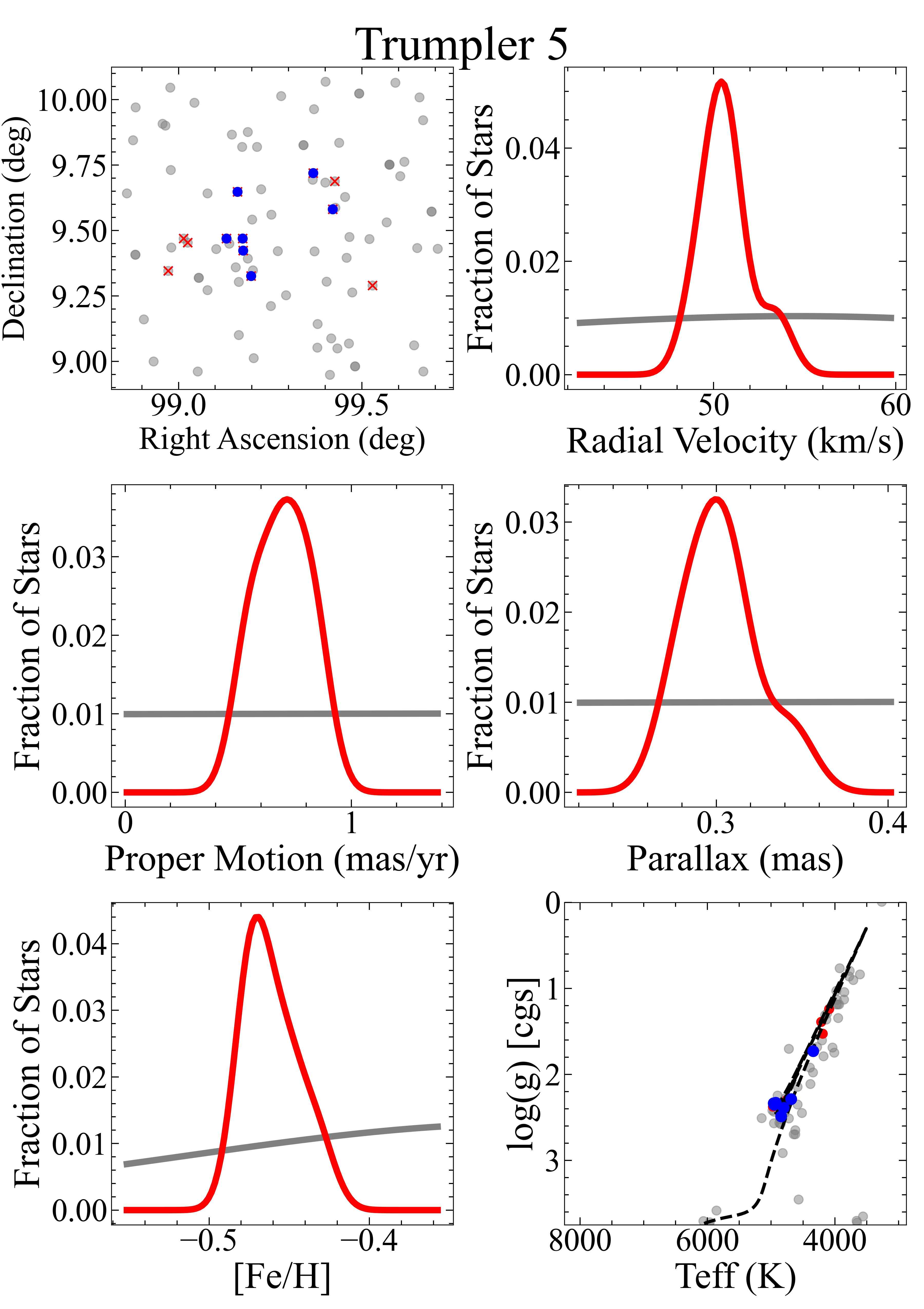}
    \caption{These figures follow the same layout as Figure \ref{fig:kinematic} in Section \ref{sec:kinematics}.}
\end{figure}
\end{document}